\documentclass[%
reprint,
superscriptaddress,
 amsmath,amssymb,
 aps,
prl,
]{revtex4-1}

\usepackage[latin9]{inputenc}
\usepackage{amsmath}
\usepackage{graphicx}

\makeatletter
 
\usepackage{epstopdf}
\usepackage{array}
\usepackage{mathrsfs}
\usepackage{dcolumn}
\usepackage{bm}

\newcolumntype{L}[1]{>{\raggedright\let\newline\\\arraybackslash\hspace{0pt}}m{#1}}
\newcolumntype{C}[1]{>{\centering\let\newline\\\arraybackslash\hspace{0pt}}m{#1}}
\newcolumntype{R}[1]{>{\raggedleft\let\newline\\\arraybackslash\hspace{0pt}}m{#1}}

\graphicspath{{../../Inkscape/}{../../Mathematica/}}

\newcommand{\BaNi}{BaNi$_2$As$_2$}

\newcommand{\BaFe}{BaFe$_2$As$_2$}
\newcommand{\BaSr}{Ba$_{1-x}$Sr$_{x}$Ni$_2$As$_2$}
\newcommand{\Tc}{$T_c$}

\newcommand{\mychi}{\raisebox{0pt}[1ex][1ex]{$\chi$}}

\usepackage{tikz}
\usepgflibrary{shadings}
\usepackage{graphicx}
\usepackage{epstopdf}
\usepackage{textcomp}
\usepackage{dcolumn}
\usepackage{bm}
\usepackage{gensymb}
\usepackage{upgreek}
\begin{document}

\preprint{APS/123-QED}

\title{Sixfold enhancement of superconductivity in a tunable electronic nematic system}

\author{Chris Eckberg}
\affiliation{Center for Nanophysics and Advanced Materials, Department of Physics, University of Maryland, College Park, Maryland 20742, USA}

\author{Daniel J. Campbell}
\affiliation{Center for Nanophysics and Advanced Materials, Department of Physics, University of Maryland, College Park, Maryland 20742, USA}
\author{Tristin Metz}
\affiliation{Center for Nanophysics and Advanced Materials, Department of Physics, University of Maryland, College Park, Maryland 20742, USA}
\author{John Collini}
\affiliation{Center for Nanophysics and Advanced Materials, Department of Physics, University of Maryland, College Park, Maryland 20742, USA}
\author{Halyna Hodovanets}
\affiliation{Center for Nanophysics and Advanced Materials, Department of Physics, University of Maryland, College Park, Maryland 20742, USA}
\author{Tyler Drye}
\affiliation{Center for Nanophysics and Advanced Materials, Department of Physics, University of Maryland, College Park, Maryland 20742, USA}
\author{Peter Zavalij}
\affiliation{Department of Chemistry, University of Maryland, College Park, Maryland 20742, USA}
\author{Morten H. Christensen}
\affiliation{School of Physics and Astronomy, University of Minnesota, Minneapolis, Minnesota 55455, USA}
\author{Rafael M. Fernandes}
\affiliation{School of Physics and Astronomy, University of Minnesota, Minneapolis, Minnesota 55455, USA}
\author{Sangjun Lee}
\affiliation{Department of Physics, Seitz Materials Research Laboratory, University of Illinois at Urbana-Champaign, Urbana, IL 61801, USA}
\author{Peter Abbamonte}
\affiliation{Department of Physics, Seitz Materials Research Laboratory, University of Illinois at Urbana-Champaign, Urbana, IL 61801, USA}
\author{Jeffrey Lynn}
\affiliation{NIST Center for Neutron Research, National Institute of Standards and Technology, Gaithersburg, Maryland 20899, USA}

\author{Johnpierre Paglione}
\email{paglione@umd.edu}
\affiliation{Center for Nanophysics and Advanced Materials, Department of Physics, University of Maryland, College Park, Maryland 20742, USA}
\affiliation{The Canadian Institute for Advanced Research, Toronto, ON M5G 1Z8, Canada}

\date{\today}

\begin{abstract}
The electronic nematic phase, wherein electronic degrees of freedom lower the crystal rotational symmetry, is a common motif across a number of high-temperature superconductors. However, understanding the role and influence of nematicity and nematic fluctuations in Cooper pairing is often complicated by the coexistence of other orders, particularly long-range magnetic order. Here we report the enhancement of superconductivity in a model electronic nematic system absent of magnetism, and show that the enhancement is directly born out of strong nematic fluctuations emanating from a tuned quantum phase transition. We use elastoresistance measurements of the \BaSr~substitution series to show that strontium substitution promotes an electronically driven $B_{1g}$ nematic order in this system, and that the complete suppression of that order to absolute zero temperature evokes a dramatic enhancement of the pairing strength, as evidenced by a sixfold increase in the superconducting transition temperature. The direct relation between enhanced pairing and nematic fluctuations in this model system, as well as the interplay with a unidirectional charge density wave order comparable to that found in the cuprates, offers a means to elucidate the role of nematicity in boosting superconductivity.
\end{abstract}

\maketitle

\section{ }

High-temperature superconductivity in both cuprate \cite{Orenstein2000,Keimer2015} and iron-based materials \cite{PaglioneRev,JohnstonRev,StewartRev}
emerges from a notably complex normal state. Though magnetic spin fluctuations are commonly believed to drive Cooper pairing in both of these families, the common occurrence of a rotational symmetry-breaking nematic phase has captured increasing attention in recent years \cite{Fernandes,Kivelson}. In contrast to a conventional structural transition, overwhelming evidence suggests that the nematic phase in these compounds is promoted by an electronic instability rather than lattice softening  \cite{ChuNematic,KuoNematic}. 

Theoretical analyses have shown that fluctuations associated with such an electronic nematic phase, particularly near a putative quantum critical point, can significantly enhance superconductivity \cite{Metlitski2015,Lederer,LedererPNAS,Fernandes_Kang,Klein2018}. Being peaked at zero wave-vector, nematic fluctuations favor pairing instabilities in several symmetry channels, in contrast to the case of magnetic fluctuations. Experiments have indeed shown a striking enhancement of nematic fluctuations centered at optimal tuning of superconductivity in a number of iron-based superconductors \cite{ChuNematic,KuoNematic}, and a strong tendency towards nematicity in high \Tc~ cuprate materials \cite{Hinkov2008,VojtaRev,Sato2017}. However, the overarching presence of magnetic fluctuations emanating from proximate antiferromagnetic instabilities complicates drawing any isolated relation between enhanced pairing and nematicity in most nematic materials. The FeSe$_{1-x}$S$_x$ substitution series is one exception, where the system exhibits both superconductivity and nematicity in the absence of magnetic order \cite{Hosoi8139}. However, in this series, the superconducting transition temperature \Tc~is virtually unaffected by tuning through the nematic quantum critical point \cite{Coldea2017,Hosoi8139}, leaving open questions about the influence of nematic fluctuations.

Here we present the discovery of electronic nematicity and evidence for nematic-fluctuation enhanced superconductivity in \BaSr, a seemingly conventional nickel-based superconductor series that is readily tunable by chemical substitution and is void of magnetic order. \BaNi, the nickel-based analog of the iron-based parent compound \BaFe, is a metallic compound that exhibits a strongly first-order structural transition from tetragonal to triclinic crystal structure at $T_S=135$ K on cooling. While magnetic order has not been found in \BaNi~to date~\cite{Ronning,Sefat,Neutron}, recent x-ray diffraction measurements have provided evidence for a previously unobserved incommensurate and unidirectional charge density wave (CDW) order that onsets at temperatures above $T_S$ \cite{Abbamonte}, followed by an abrupt transition to a new, commensurate CDW order upon cooling into the triclinic phase \cite{Abbamonte}. In contrast, in the 100\% Sr end-member SrNi$_2$As$_2$ a tetragonal phase persists down to absolute zero temperature and no known CDW or magnetic orders exist \cite{Bauer}. Both materials superconduct near $0.6$ K, and thermodynamic experiments as well as first-principle analysis have indicated that superconductivity is of conventional origin in both materials \cite{Subedi,Kurita}. In this work, we study the superconducting phase as a function of Ba/Sr content, finding that it is not only robust but is dramatically enhanced when electronic nematic fluctuations are maximized at a quantum phase transition.

\begin{figure}
    \centering
    \includegraphics[width=0.47\textwidth]{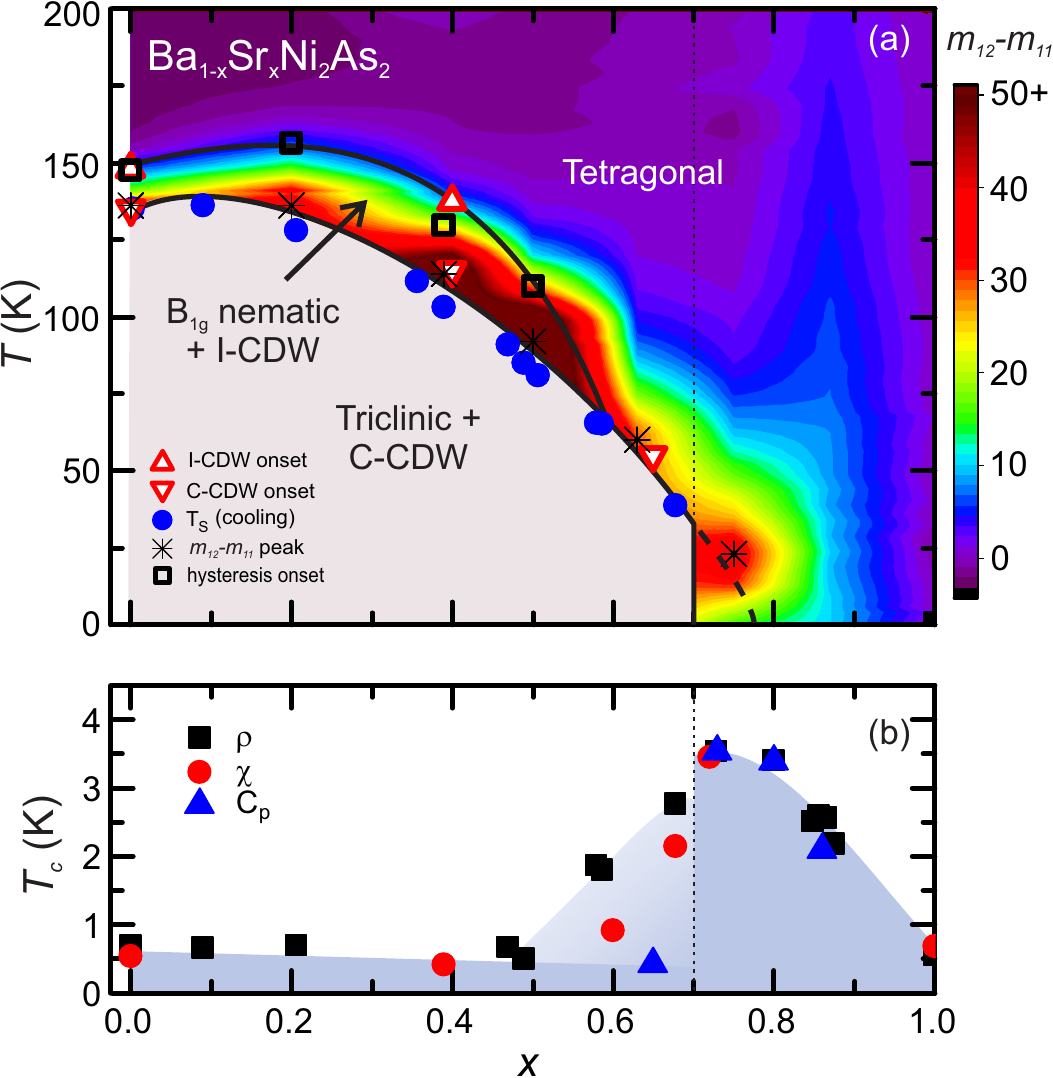}
    \caption{{\bf Evolution of structural, charge and nematic orders in \BaSr.} 
    This system presents a rich interplay of structural, charge and nematic instabilities that evolve as a function of chemical pressure induced by Sr substitution. It features a dramatic enhancement of the superconducting transition temperature in the region where charge and nematic orders cease to be long-range, and nematic fluctuations are peaked at the lowest temperatures. (a) Single-crystal measurements form a phase diagram consisting of onsets of incommensurate charge-density wave (I-CDW) order (upright red triangles), elastoresistive strain-hysteresis (open black squares), commensurate charge-density wave (C-CDW) order (inverted red triangles), and the cooling transition of the first-order triclinic structural distortion (closed blue circles). Black asterisks mark the peak position of nematic susceptibilities, which extend beyond the disappearance (vertical dashed line) of the triclinic phase at $x_c=0.70$. The overlayed color scale represents interpolated values of the nematic susceptibility $m_{12}-m_{11}$ generated from data taken in \BaSr~single crystals with $x=0$, 0.2, 0.4, 0.5, 0.63, 0.75, 0.87, and 1.0. 
    (b) Superconducting transition temperatures $T_c$ in \BaSr~single crystals determined by transport (black squares), magnetization (red circles), and heat capacity (blue triangles) measurements. Dark blue shading reflects regions of bulk superconductivity (as confirmed via heat capacity, magnetization and transport transitions) while the light blue region indicates filamentary superconductivity observed as broad transitions in transport and magnetic measurements, but absent in heat capacity.
    }
    \label{fig:Figure3}
\end{figure}

\begin{figure}
    \centering
    \includegraphics[width=0.45\textwidth]{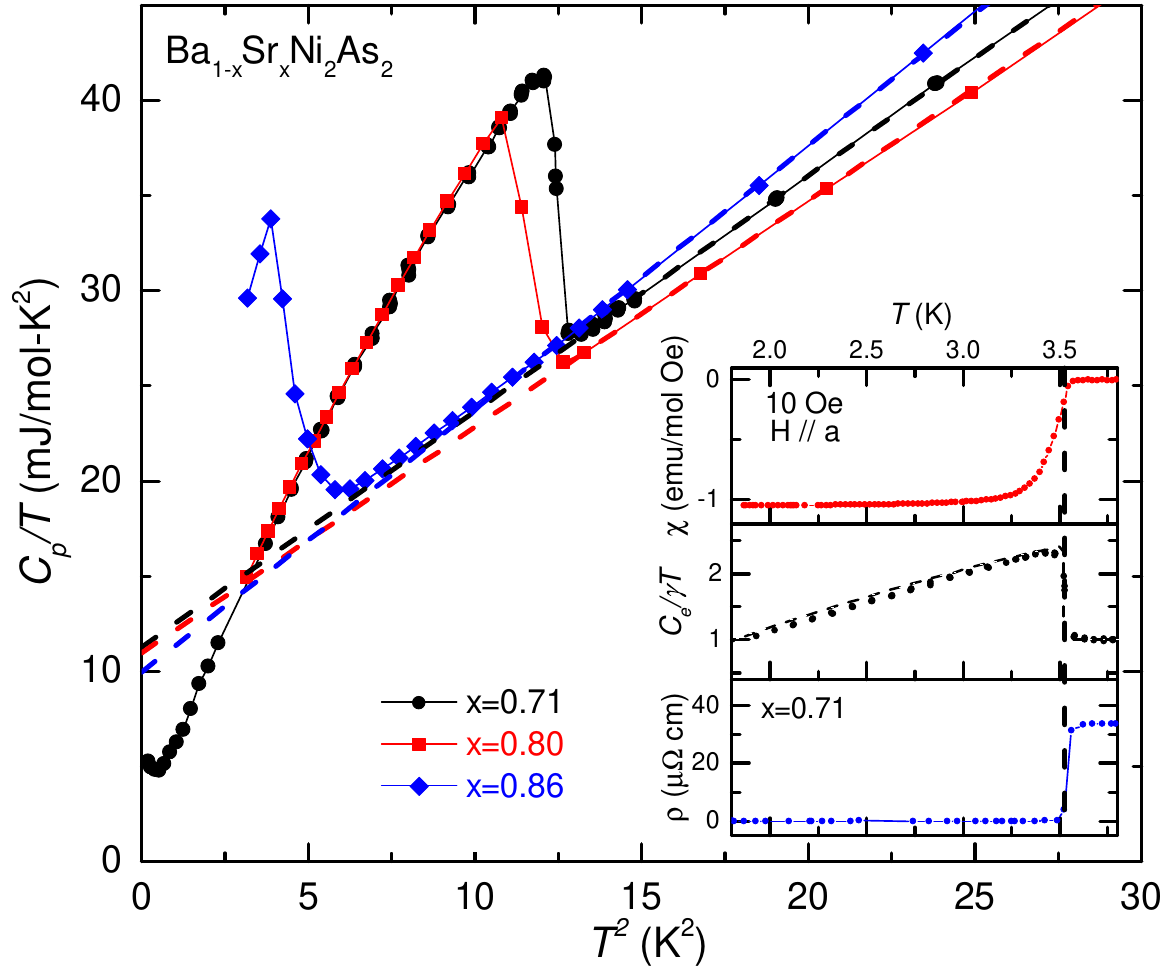}
    \caption{(a) {\bf Enhancement of superconducting transition temperature.}
    Heat capacity measurements in slightly over-substituted \BaSr~single crystals with $x=0.71,~0.80$, and 0.86 depict dramatic enhancement of $T_c$. This contrasts with the very small changes in Sr concentrations, Debye temperatures (as determined by the $T^3$ phonon contribution shown by the dashed lines), and density of states (as determined by Sommerfeld coefficients, given by the extrapolation of the dashed lines to $T=0$) across these samples. 
    (inset) The superconducting transition in the same $x=0.71$ single-crystal specimen, measured by magnetization (upper panel), electronic heat capacity (center panel), and transport (lower panel) are consistent with a strongly enhanced $T_c$ of 3.5 K, far above the values at either series end member. The dashed line in (c) indicates the predicted electronic heat capacity anomaly for a single band, s-wave superconductor, with a BCS gap \cite{Johnston2013}.}
    \label{fig:Figure4}
\end{figure}
\begin{figure}
    \centering
    \includegraphics[width=0.47\textwidth]{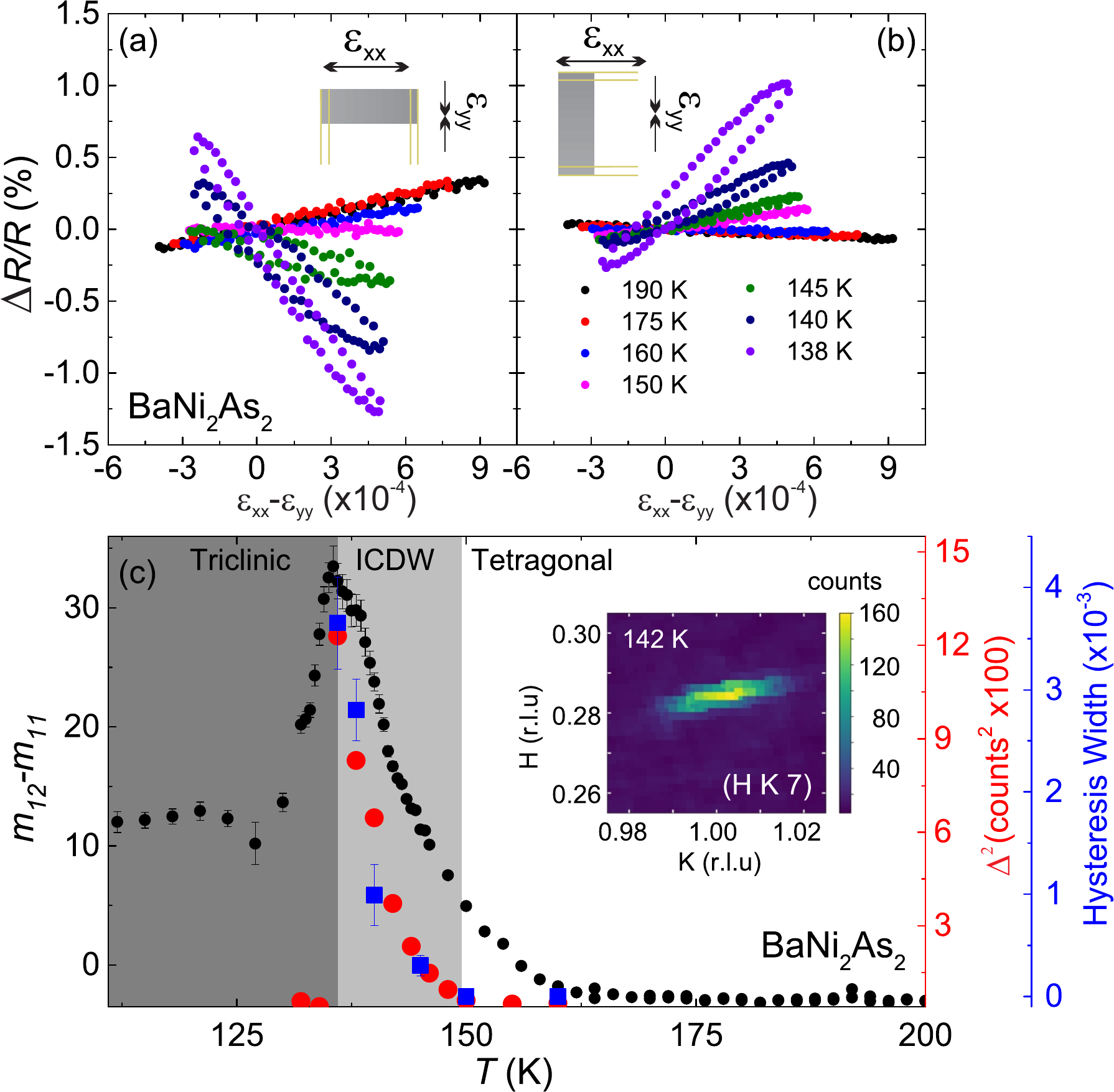}
    \caption{{\bf Electronic nematic and charge orders in \BaNi.}
     A strongly divergent $B_{1g}$ nematic susceptibility, determined by elastoresistivity measurements shown in the strain-dependent resistance isotherms for single crystals mounted parallel (a) and perpendicular (b) to the poling direction of the piezo stack, is comparable in magnitude to that of \BaFe~\cite{ChuNematic} and is accompanied by two notable ordered states. 
     As shown in panel (c), the nematic susceptibility in the $B_{1g}$ channel, proportional to the elastoresistance $m_{12}-m_{11}$, is nearly flat at high temperatures before growing upon approaching the incommensurate charge-density wave (I-CDW) ordered phase (light grey region), and then peaking at the structural transition into the triclinic phase (dark grey region). (Black symbols include error bars representing 90\% confidence intervals of data). Strain-hysteretic behavior of the elastoresistance hysteresis is observed to onset at the same temperature where I-CDW is seen.
     A comparison of the squared peak intensity of a $(0.28~0~0)$ I-CDW superstructure reflection [the $(-1.72~1~7)$ peak; red symbols] and the elastoresistive hysteresis width (blue symbols), show a nearly linear relationship. Strain-dependent isotherms were repeated three times at each temperature, and hysteresis widths were measured at the widest point. Error bars represent extremal values of the hysteresis width between separate measurements. Inset shows an $(H~K)$ map of the reciprocal space at 142 K, displaying a reflection from the superstructure at wave vector $(0.28~1~7)$.}
    \label{fig:Figure1}
\end{figure}

\begin{figure*}
    \centering
    \includegraphics[width=1\textwidth]{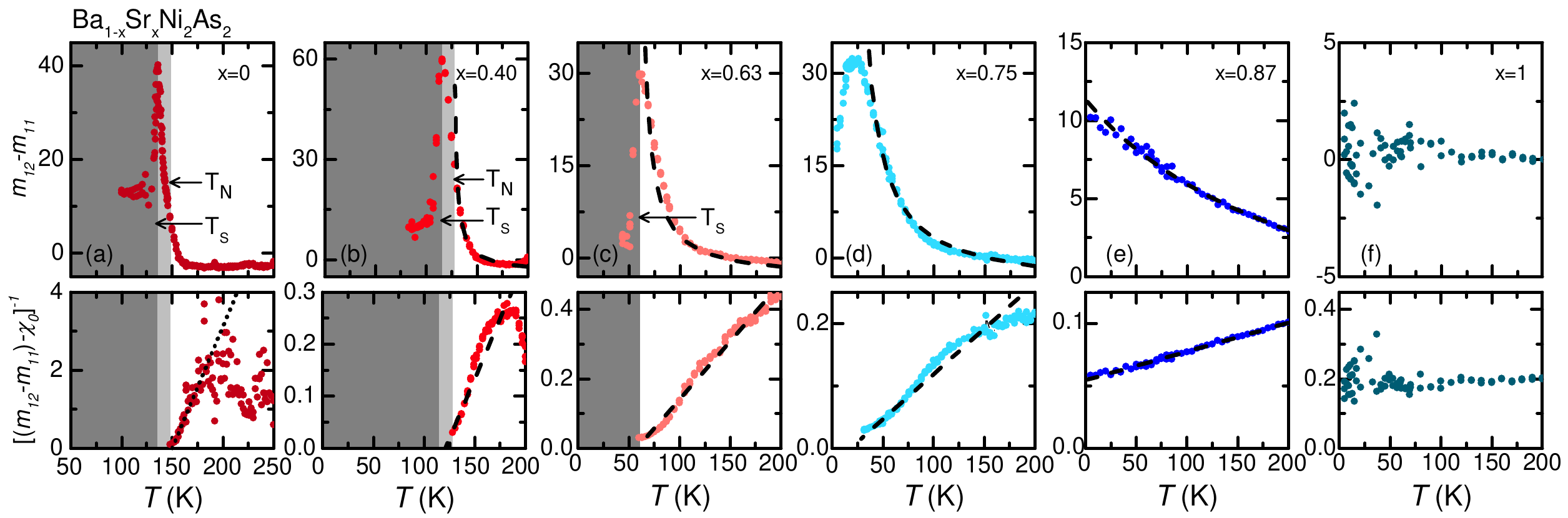}
    \caption{{\bf Nematic susceptibilities of \BaSr~single crystals.}
    The $B_{1g}$ nematic susceptibility exhibits a continuous evolution with $x$, eventually disappearing in SrNi$_2$As$_2$ ($x$=1), as shown in the top panels. 
    Dark grey regions indicate triclinic structural phases in $x=0$, 0.40 and 0.63, and light grey regions (only in $x=0$ and 0.40) indicate temperatures where elastoresistive hysteresis is observed in the tetragonal structure. 
    Lower panels display the inverse susceptibilities $[(m_{12}-m_{11})-\chi_0]^{-1}$, truncated at the onset of nematic order. The constant, $\chi_0$, is a temperature independent component of the elastoresistance, coming from factors unrelated to nematic order, including changing sample geometry, and is determined through fitting data to the modified Curie-Weiss functional form; $m_{12}-m_{11}=\frac{\lambda}{a_0(T-T_\mathcal{N})}+\chi_0$. Black dashed lines show results of this fitting. No fitting is presented in samples of $x=0$ or $x=1$, as neither of these show Curie-Weiss like divergence. The dotted black line in the lower panel of (a) is rather a guide for the eye, indicating incipient nematic fluctuations in the vicinity of $T_\mathcal{N}$. At $x=0.75$, which exhibits no nematic, structural or charge ordered transitions and remains tetragonal to the lowest temperatures, the nematic susceptibility exhibits a broad peak at 25~K. }
    \label{fig:Figure2}
\end{figure*}

We begin by presenting the global phase diagram of the \BaSr~system in Fig. 1(a). Replacing Sr for Ba suppresses the first-order tetragonal-triclinic structural distortion, as well as the simultaneous commensurate CDW (C-CDW) order, in a continuous manner. Structural and C-CDW transition temperatures remain pinned to each other, but decrease in temperature as a function of $x$ until they are suppressed to $T$=0 at a critical Sr concentration of $x_c=0.70$. The incommensurate CDW (I-CDW) phase, which onsets above $T_S$ and is denoted by upright red triangles in Fig. 1, is also suppressed with increasing $x$ until it merges with $T_S$ and disappears altogether. The coherence length of the I-CDW phase, estimated from full width at half maximum of its associated x-ray peaks, is reduced from over 1000~\AA~to only several hundred \AA~between $x=0$ and $x=0.4$, and is entirely absent in samples with larger $x$. For instance, at $x=0.65$ the system transitions from a tetragonal structure with no charge-order superstructure peaks into a triclinic structure with commensurate C-CDW superstructure at 50~K. With increasing Sr concentration, the system exhibits an abrupt 0 K tetragonal-triclinic structural phase transition at critical value $x_c=0.7$ where a predisposition to As-As bonding reconfiguration may occur (see SI), and remains void of any charge order between $x_c$ and $x=1$.

In parallel, superconductivity in \BaSr~single crystals evolves dramatically across the phase diagram [Fig. 1(b)]. Approaching $x_c$ from the \BaNi~end member, resistive and magnetic signatures of \Tc~appear to climb in temperature at concentrations above $x=0.5$ before reaching a maximum centered just above $x_c$. Interestingly, the ``bulk'' signature of \Tc, as determined by specific heat measurements (c.f. Fig.~2), remains relatively constant versus $x$ until it abruptly jumps upon crossing the zero-temperature structural phase boundary, from approximately $0.5$ K in $x=0.68$ samples to near $3.5$ K at $x=0.71$. Optimally substituted $x=0.71$ presents a very robust superconducting transition as measured by resistivity, diamagnetism, and specific heat anomaly, well fit by a single band model with a BCS gap [Fig.~2 inset].
Remarkably, this optimal \Tc~of 3.5 K at $x=0.71$ marks a nearly sixfold enhancement compared to either of the Sr- or Ba-based end members. The superconducting transition then decreases continuously with increasing $x$ toward that of SrNi$_2$As$_2$, all through a regime with no notable change in structure or appearance of charge-ordered phases. Moreover, this dramatic change in pairing strength occurs through a regime with virtually unchanged Sommerfeld coefficients of approximately 10 mJ/mol-K$^2$ [Fig. 2 main], indicating that the \Tc~enhancement cannot be explained by changes in the electronic density of states.

Consistent with prior studies \cite{KudoP,KudoCu,Eckberg}, the \BaSr~system exhibits a discontinuity in the Debye temperature $\Theta_D$ at $x_c$. However, $\Theta_D$ remains approximately constant between $x=0.71$ ($\Theta_D$=198 K) and $x=0.86$ ($\Theta_D$=188 K), despite a nearly two-fold difference in the superconducting \Tc. This contrast indicates that, like the electronic density of states, changes in lattice stiffness do not come close to capturing the enhancement in the pairing potential. Rather, the smooth increase in critical temperature upon approaching $x_c$ from above is reminiscent of a fluctuation-driven superconducting enhancement. Given the very abrupt first-order nature of the triclinic-tetragonal structural boundary, which appears to drop precipitously at $x_c$, an increase in pairing strength must arise from a hitherto hidden coupling to the electronic system that can enhance pairing.

Having ruled out the usual sources of a \Tc~increase expected for a conventional phonon-mediated superconductor (density of states and Debye frequency), we propose that this enhancement is driven by the presence of nematic fluctuations. Indeed, the existence of a tetragonal-to-triclinic transition shows that rotational symmetry is broken in this system. To investigate whether this transition is driven by electronic, rather than lattice degrees of freedom, we perform elastoresistance measurements. The elastoresistance tensor $m_{ij,kl}$ corresponds to the rate of change of the normalized resistivity $(\Delta\rho/\rho)_{ij}$ upon application of external strain $\epsilon_{kl}$, i.e. $m_{ij,kl} = \frac{\partial (\Delta\rho/\rho)_{ij}}{\partial \epsilon_{kl}}$. As discussed in \cite{ChuNematic,ShapiroTransverse}, when the applied strain transforms as one of the non-trivial irreducible representations $\Gamma_\mu$ of the point group, $\epsilon_{kl} \equiv \varepsilon_\mu$, the resistivity change in that channel is proportional to the corresponding nematic order $\psi_\mu$, $(\Delta\rho/\rho)_{ij} \propto \psi_\mu$. Because $\varepsilon_\mu$ acts as a conjugate field to the nematic order parameter $\psi_\mu$, the corresponding elastoresistance $m_{\Gamma_\mu}$ becomes proportional to the electronic contribution to the nematic susceptibility, i.e. the bare nematic susceptibility without renormalization by the lattice degrees of freedom:  

\begin{equation}
m_{\Gamma_\mu} \propto \mychi_{\mathrm{nem}}^{\mu} \equiv \frac{\partial \psi_\mu}{\partial \varepsilon_\mu} \label{eq:free_energy}
\end{equation}  

The key point is that, if $m_{\Gamma_\mu}$ shows a diverging behavior above the rotational symmetry breaking transition, it implies that the latter is nematic, i.e. driven by electronic degrees of freedom. If $m_{\Gamma_\mu}$ shows instead a weak temperature dependence, it implies that the transition is a standard lattice-driven structural transition.

In the case of tetragonal \BaSr, there are three symmetry-distinct channels of rotational symmetry breaking, corresponding to the three irreducible representations $B_{1g}$, $B_{2g}$, and $E_g$ of the point group $D_{4h}$. In terms of charge degrees of freedom, they correspond to quadrupolar charge order with form factors $x^2-y^2$, $xy$, and $\left( xz, yz\right)$, respectively. In terms of lattice degrees of freedom, the first two correspond to orthorhombic distortions $\varepsilon_{B_{1g}}$ and $\varepsilon_{B_{2g}}$, and the third, to a monoclinic distortion $\left( \varepsilon_{E_{g}}^{1},\varepsilon_{E_{g}}^{2} \right)$ of the tetragonal lattice. Importantly, in the triclinic phase of \BaSr , all four lattice distortions are present. This indicates the potential that one or more of the three nematic susceptibilities $\mychi_{\mathrm{nem}}^{B_{1g}}$, $\mychi_{\mathrm{nem}}^{B_{2g}}$, and $\mychi_{\mathrm{nem}}^{E_{g}}$ may be diverging above $T_S$.

To measure the nematic susceptibilities, we use the piezoelectric elastoresistance technique of Refs. \cite{ChuNematic,ShapiroTransverse}, applying in-situ tunable biaxial strain to \BaSr~single crystal specimens. Figure 3 presents the elastoresistance $m_{12}-m_{11}$, which is proportional to $\mychi_{\mathrm{nem}}^{B_{1g}}$, in stoichiometric \BaNi. While $m_{12}-m_{11}$ is negative at temperatures well above the structural transition, it becomes positive near $T_S\approx 135$ K [Fig. 3(a,b)]. Before it peaks at $T_S$, however, $m_{12}-m_{11}$ starts displaying strain-hysteretic behavior at a temperature of about $148$ K [blue symbols in Fig. 3(c)]. This temperature coincides with the emergence of CDW peaks at the incommensurate wave vector $(0.28~0~0)$ [red symbols in Fig. 3(c)]. Significantly, the CDW peaks corresponding to this structure are observed across many Brillouin zones, while no peaks are seen in the orthogonal $(0~0.28~0)$ orientation. While $m_{12}-m_{11}$ data are still presented in the temperature range of strain hysteretic resistance in Fig. 3(c), $T<148$ K, it is important to note that these values are no longer true nematic susceptibilities, since the latter is only well-defined in the regime of linear response. The observation of unidirectional CDW peaks and a strain-hysteretic $m_{12}-m_{11}$ in \BaNi~indicate a static tetragonal symmetry-breaking phase in the $B_{1g}$ channel at a temperature $T_\mathcal{N}$ that is higher than $T_S$. Crucially, the electronic nematic susceptibility, proportional to $m_{12}-m_{11}$, does not seem to diverge near $T_\mathcal{N}$ -- in fact, it is nearly temperature-independent above  $T_\mathcal{N}$. This is indicative that the transition is driven primarily not by electronic, but by lattice degrees of freedom.

The evolution of the elastoresistivity in the $B_{1g}$ channel in doped \BaSr~crystals is presented in Fig. 4. First, we note that the onset of strain-hysteretic behavior at $T_{\mathcal{N}}$, indicated by the light-gray shaded areas in the plots, moves closer to the triclinic structural transition $T_S$, and eventually merge with the latter for $x=0.63$. Second, the modest temperature dependence of $m_{12}-m_{11}$ above $T_\mathcal{N}$ in stoichiometric \BaNi~is not reflected in more heavily-substituted samples. Indeed, $m_{12}-m_{11}$ starts displaying a diverging behavior above $T_\mathcal{N}$ over a wide temperature range in \BaSr~samples with increasing $x$. These data may be reasonably fit to a modified Curie-Weiss function ($\mychi_{\mathrm{nem}}^{B_{1g}}=\frac{\lambda}{a_0(T-T_\mathcal{N})}+\chi_0$) above $T_\mathcal{N}$, indicating diverging susceptibilities reminiscent of electronically driven nematic order. Therefore, our elastoresistance data shows a change in the character of the tetragonal symmetry-breaking transition from lattice-driven for small $x$ to electronically-driven for $x$ near optimal doping. This is corroborated by a phenomenological Ginzburg-Landau calculation to model the nematic susceptibility data (see SI).

While the $B_{1g}$ nematic susceptibility diverges, the $B_{2g}$ susceptibility is only very weakly temperature dependent in samples with $x=0$ and $x=0.63$ (See SI). Unfortunately, our sample geometry does not allow for measurements of the $E_g$ nematic susceptibility. The absence of a diverging $B_{2g}$ susceptibility indicates that the strengthening of the electronic nematic fluctuations is limited to the $B_{1g}$ symmetry channel. This contrasts with the structurally related Fe-based superconductors, where ubiquitous signatures of nematicity in the $B_{2g}$ channel are reported~\cite{KuoNematic}. Such a lack of enhancement happens in spite of the fact that rotational symmetry-breaking in the $B_{2g}$ channel happens at the triclinic transition. This indicates that the triclinic phase transition cannot be attributed solely to electronic degrees of freedom.

The diverging $B_{1g}$ susceptibility persists in the $x=0.75$ samples, which feature no discernible phase transition in thermal, magnetic, transport, or diffraction measurements down to the lowest temperatures. Despite the absence of any evident phase transition, $m_{12}-m_{11}$ data for Ba$_{0.25}$Sr$_{0.75}$Ni$_2$As$_2$ exhibit a clear peak and subsequent downturn at 25~K. Such a nematic susceptibility peak, in the absence of any apparent order (see SI), is unprecedented in its observation. In analogy to more familiar magnetic systems, it may be an indication of a freezing {\it nematic glass}, or possibly an artifact of quenched disorder subverting long-range nematic correlations.     

Returning to the overall phase diagram of \BaSr, the amplitude of the $B_{1g}$ nematic susceptibility is overlaid with the triclinc and CDW phase boundaries in Fig. 1. Beginning from the SrNi$_2$As$_2$ end member, a smooth enhancement of electronic nematic fluctuations is readily observed at the lowest temperatures. In stark contrast to the stagnant behavior of other thermodynamic quantities, such as specific heat, dramatic strengthening nematic fluctuations grow concurrently with rapidly enhancing superconducting $T_c$, with an over ten-fold enhancement of $m_{12}-m_{11}$ from $x=1$ to $x_c$. It is through the exchange of these excitations that the superconducting enhancement can be explained \cite{Lederer, LedererPNAS, Fernandes_Kang, Klein2018}. The strength of this enhancement, corresponding to a nearly sixfold increase in \Tc~from the series end members, establishes nematic fluctuations as a promising mechanism for dramatically enhancing Cooper pairing, even in a conventional superconductor such as, presumably, the one studied here.

The origin of nematic order cannot be inferred solely from elastoresistance measurements. In the structurally and chemically similar \BaFe~compounds, the $B_{2g}$ electronic nematic order is proposed to be driven by magnetic degrees of freedom, since the stripe magnetic ground state breaks the tetragonal symmetry in the same channel \cite{Cano2010,Fernandes2012,Fernandes_Bohmer}. The \BaSr~series, in contrast, exhibits no known magnetic order. However, it does exhibit unidirectional CDW order that also breaks tetragonal symmetry in the $B_{1g}$ channel. It is therefore tempting to attribute the nematic instability in \BaSr~as driven by charge fluctuations. Indeed, comparing in Fig. 3(c) the square of the CDW x-ray peak intensity to the width of the elastoresistance hysteresis, which is a proxy of the nematic order parameter (see SI), we observe a nearly linear relationship between the two quantities, as expected by symmetry considerations. This lends further support to the assumption of charge-driven nematicity. It also provides a compelling scenario to explain the phase diagram of \BaSr~in terms of two cooperative ordered states: a charge-driven electronic nematic phase and a lattice-driven triclinic phase. While both break the tetragonal symmetry in the $B_{1g}$ channel, the latter also breaks additional symmetries that our elastoresistance measurements show cannot be accounted for solely by electronic degrees of freedom (as evidenced by the lack of divergence of the $B_{2g}$ nematic susceptibility).

The likely relationship between nematic and CDW order evokes comparison to the cuprate superconductors, wherein short range, unidirectional, incommensurate CDW stripe order and electronic anisotropies have been reported in the pseudogap phase \cite{Kivelson,Achkar,VojtaRev}. In the cuprates, it has been proposed that the microscopic tendency is towards unidirectional CDW order, with long range coherence being precluded by quenched disorder \cite{Nie}. The nematic phase is more robust to disorder, however, surviving as a vestige of the suppressed stripes. In \BaSr, unlike the cuprate compounds, long range CDW superstructures survive for sufficiently small~$x$. 

\section{\label{sec:Methods}Methods}

\subsection{Crystal Synthesis}

\BaSr~single crystals were synthesized using prereacted NiAs self flux combined with Ba and Sr pieces in a 4:$1-x$:$x$ ratio as previously reported \cite{Sefat}. Materials were  heated to 1180\degree C before being slowly cooled to 980\degree C at 2 degrees an hour. At this point the furnace was turned off and allowed to cool to room temperature naturally. Once cool, crystals with typical dimensions of 2mm x 2mm x 0.5mm were mechanically extracted from flux. Chemical compositions of resulting crystals were determined using a combination of energy dispersive spectroscopy (EDS) and single crystal x-ray refinements.

\subsection{Transport, Specific Heat, and Magnetization}

Transport and heat capacity data were taken using both a Quantum Design Physical Property Measurement System and a Quantum Design DynaCool. Heat capacity data were generally collected via a relaxation method. To observe the first order phase transition upon both warming and cooling, select heat capacity measurements were modified to be sensitive to both transitions. Within these modified measurements, an extended heat pulse was applied and heat capacity was extracted using a local derivative approach. Dc-magnetization measurements were taken using a SQUID-VSM option in a Quantum Design MPMS3 system. A homemade coil \cite{Coil} was also used in a Quantum Design adiabatic demagnetization refrigerator (ADR) insert to measure ac-susceptibility down to 0.1 K.

\subsection{X-ray Diffraction}

250 K structural data were collected on single crystals in a Bruker APEX-II CCD system equipped with a graphite monochromator and a MoK$_\alpha$~sealed tube ($\lambda$ = 0.71073 \AA), and were refined using the Bruker SHELXTL Software Package. Temperature dependent diffraction measurements were carried out using a Xenocs GeniX 3D MoK$_\alpha$ microspot x-ray source with multilayer focusing optics and a Mar345 image plate detector. Single crystal samples were cooled with a closed-cycle cyrostat and mounted to a Huber four-circle diffractometer.

\subsection{Elastoresistivity}

As a measure of the thermodynamic nematic susceptibility, elastoresistive measurements were taken across the \BaSr~phase diagram. Within the $D_{4h}$ point group, elastoresistive coefficient $m_{12}-m_{11}$ are directly proportional to $\mychi_{\mathrm{nem}}^{B_{1g}}$ while $m_{66}$ is directly proportional to $\mychi_{\mathrm{nem}}^{B_{2g}}$ \cite{ShapiroTensor}. Both $m_{12}-m_{11}$ and $m_{66}$ were measured in samples adhered directly to a lead-zirconium-titanate (PZT) piezoelectric stack using a strain transmitting epoxy as discussed in Refs. \cite{ChuNematic,ShapiroTransverse}. By applying a voltage to the stack, variable bi-axial strain was applied in-situ. The magnitude of the applied strain was measured using a strain gauge mounted on the reverse side of the stack. The strain was measured along a single piezo axis ($\epsilon_{xx}$ in the convention used within this text), and orthogonal strain was calculated using the known Poisson's ratio of the stack. $m_{12}-m_{11}$ elastoresistive coefficients were measured using two samples mounted in a mutual orthogonal geometry, with crystal $(1~0~0)$ axis mounted parallel and perpendicular to the piezo poling direction. For all measurements requiring two samples, a single crystal was polished to a suitable thickness ($\leq$60$\mu m$) and then cleaved into two pieces. These two pieces were used for a single measurement in order to assure consistent physical properties across the two orthogonal samples. $m_{66}$ data were collected using a single sample wired in a transverse geometry. Strain was then applied along the crystallographic $(1~1~0)$ direction \cite{ShapiroTransverse}. Crystal geometry [specifically a narrow c-axis cross section and tendency towards cleaving along the crystallographic $(1~0~0)$ direction] made $m_{44}$, $\mychi_{\mathrm{nem}}^{E_{g}}$, experimentally inaccessible.

\section{\label{sec:Acknowledge}Acknowledgments}

Research at the University of Maryland was supported by the AFOSR Grant No. FA9550-14-10332 and the Gordon and Betty Moore Foundation Grant No. GBMF4419. We also acknowledge support from the Center for Nanophysics and Advanced Materials as well as the Maryland Nanocenter and its FabLab. The identification of any commercial product or trade name does not imply endorsement or recommendation by the National Institute of Standards and Technology. Theory work (RMF and MHC) was supported by the U.S. Department of Energy, Office of Science, Basic Energy Sciences under award number DE-SC0012336. X-ray experiments at UIUC were supported by DOE grant DE-FG02-06ER46285. P.A. acknowledges support from the Gordon and Betty Moore Foundation's EPiQS initiative through grants GBMF4542.


\bibliography{main}

\pagebreak
\clearpage
\widetext

\begin{center}
\textbf{\large Supplemental Information: Sixfold enhancement of superconductivity in a tunable electronic nematic system}
\end{center}
\setcounter{equation}{0}
\setcounter{figure}{0}
\setcounter{table}{0}
\setcounter{section}{0}
\makeatletter
\renewcommand{\theequation}{S\arabic{equation}}
\renewcommand{\thefigure}{S\arabic{figure}}

\section{Phenomenological model}

To model the nematic and triclinic phase transitions in Ba$_{1-x}$Sr$_{x}$Ni$_{2}$As$_{2}$,
we use a phenomenological Landau free-energy expansion. We first focus
on the elastic contribution to the free energy. Since BaNi$_{2}$As$_{2}$
has a tetragonal crystal structure at high temperatures, the harmonic part of the elastic
free energy is written, in the compactified Voigt notation, as~\cite{cowley76}
\begin{eqnarray}
\mathcal{F}_{\mathrm{elast}} & = & \sum_{ij}^{6}\varepsilon_{i}C_{ij}\varepsilon_{j}\,,
\end{eqnarray}
Here, the strain fields $\varepsilon_{i}$ are 
\begin{eqnarray}
\varepsilon_{1} & = & \partial_{x}u_{x}+\partial_{y}u_{y}+\partial_{z}u_{z}\,,\\
\varepsilon_{2} & = & \frac{1}{6}\left(\partial_{x}u_{x}+\partial_{y}u_{y}-2\partial_{z}u_{z}\right)\,,\\
\varepsilon_{3} & = & \frac{1}{\sqrt{2}}\left(\partial_{x}u_{x}-\partial_{y}u_{y}\right)\,,\\
\varepsilon_{4} & = & \frac{1}{2}\left(\partial_{y}u_{z}+\partial_{z}u_{y}\right)\,,\\
\varepsilon_{5} & = & \frac{1}{2}\left(\partial_{x}u_{z}+\partial_{z}u_{x}\right)\,,\\
\varepsilon_{6} & = & \frac{1}{2}\left(\partial_{x}u_{y}+\partial_{y}u_{x}\right)\,,
\end{eqnarray}
where $u_{i}$ are the components of the displacement vector and $C_{ij}$
is a matrix containing the elastic moduli. Here we shall focus on
the modes that lead to some form of rotational symmetry breaking,
which correspond to $\varepsilon_{3}$, $\varepsilon_{4}$, $\varepsilon_{5}$,
and $\varepsilon_{6}$. $\varepsilon_{1}$ and $\varepsilon_{2}$
correspond to structural deformations that do not break the tetragonal
symmetry. In the language of irreducible representations (irreps)
of the tetragonal group, $\varepsilon_{3}$ transforms as $B_{1g}$,
$\varepsilon_{6}$ as $B_{2g}$ and $\varepsilon_{4}$ and $\varepsilon_{5}$
constitute the two components of an $E_{g}$ irrep. The condensation
of either $\mu_{3}$ or $\mu_{6}$ leads to orthorhombic crystal structures,
while $\varepsilon_{4}$ and/or $\varepsilon_{5}$ result in a monoclinic
structure. On the other hand, a triclinic structure only occurs if
all four order parameters, $\varepsilon_{3}$, $\varepsilon_{4}$,
$\varepsilon_{5}$, and $\varepsilon_{6}$, are non-zero.

Focusing on these terms and going beyond the harmonic approximation,
the elastic free energy becomes:
\begin{align}
\mathcal{F}_{\mathrm{elast}} & =\frac{1}{2}\left(C_{11}-C_{12}\right)\varepsilon_{3}^{2}+\frac{1}{2}C_{66}\varepsilon_{6}^{2}+\frac{1}{2}C_{44}\left(\varepsilon_{4}^{2}+\varepsilon_{5}^{2}\right)+\lambda_{2}\varepsilon_{4}\varepsilon_{5}\varepsilon_{6}+\lambda_{3}\left(\varepsilon_{4}^{2}-\varepsilon_{5}^{2}\right)\varepsilon_{3}\nonumber+ \\
 & \frac{u_{1}}{4}\varepsilon_{3}^{4}+\frac{u_{2}}{4}\varepsilon_{6}^{4}+\frac{u_{3}}{4}\left(\varepsilon_{4}^{2}+\varepsilon_{5}^{2}\right)^{2}-\frac{g_{3}}{4}\left(\varepsilon_{4}^{2}-\varepsilon_{5}^{2}\right)^{2},\label{F_el}
\end{align}
where we have neglected quartic terms that will not be important for the analysis below. Note that the cubic terms imply that the condensation of $\varepsilon_{4}$
and/or $\varepsilon_{5}$ necessarily leads to a non-zero $\varepsilon_{3}$
and/or $\varepsilon_{6}$, since $\varepsilon_{6}\sim\varepsilon_{4}\varepsilon_{5}$
and $\varepsilon_{3}\sim\varepsilon_{4}^{2}-\varepsilon_{5}^{2}$.
Among the quartic terms, $g_{3}>0$ favors condensation of either
$\varepsilon_{4}$ or $\varepsilon_{5}$, whereas $g_{3}<0$ favors
the simultaneous condensation of both $\varepsilon_{4}$ and $\varepsilon_{5}$.
Because in the triclinic phase $\varepsilon_{3},\varepsilon_{4},\varepsilon_{5},\varepsilon_{6}\neq0$,
a direct tetragonal-to-triclinic transition can only be first-order.
As explained in the main text, our data on undoped and underdoped
Ba$_{1-x}$Sr$_{x}$Ni$_{2}$As$_{2}$ suggests symmetry breaking
in the $B_{1g}$ channel (i.e. $\varepsilon_{3}\neq0$) before the
onset of a triclinic phase. Thus, the transition from tetragonal to
triclinic seems to happen via an intermediate orthorhombic phase,
at least for sufficiently underdoped compositions. 

To understand this sequence of transitions, consider the condensation
of a finite $\langle\varepsilon_{3}\rangle\neq0$ (either due to the
softening of the $C_{11}-C_{12}$ elastic constant or as a consequence
of an electronic nematic transition) and integrate out the $\varepsilon_{6}$
degrees of freedom. The elastic free energy becomes:
\[
\mathcal{F}_{\mathrm{elast}}=\frac{1}{2}\left(C_{44}+\lambda_{3}\langle\varepsilon_{3}\rangle\right)\varepsilon_{4}^{2}+\frac{1}{2}\left(C_{44}-\lambda_{3}\langle\varepsilon_{3}\rangle\right)\varepsilon_{5}^{2}+\frac{\left(u_{3}-\frac{\lambda_{2}^{2}}{2C_{66}}\right)}{4}\left(\varepsilon_{4}^{2}+\varepsilon_{5}^{2}\right)^{2}-\frac{\left(g_{3}-\frac{\lambda_{2}^{2}}{2C_{66}}\right)}{4}\left(\varepsilon_{4}^{2}-\varepsilon_{5}^{2}\right)^{2}.
\]
A triclinic phase transition can then take place when both $C_{44}$
and $C_{66}$ are soft enough. Indeed, when the elastic constant $C_{44}$
is small enough such that $C_{44}<\lambda_{3}|\langle\varepsilon_{3}\rangle|$,
either $\mu_{4}$ or $\mu_{5}$ will condense depending on the signs
of $\langle\varepsilon_{3}\rangle$ and $\lambda_{3}$. When $C_{66}$
is also soft enough, such that $C_{66}<\lambda_{2}^{2}/2g_{3}$, or
if $g_{3}$ is itself negative, the quartic term favors the simultaneous
appearance of both $\varepsilon_{4}$ and $\varepsilon_{5}$, which
in turn generates $\varepsilon_{6}$.
This is one possible scenario for the transition from tetragonal to
triclinic. Whether it occurs via an intermediate orthorhombic (i.e.
$\langle\varepsilon_{3}\rangle\neq0$) phase taking place at a temperature
$T_{\mathcal{N}}$ above the onset of the triclinic order at $T_{S}$
depends on the character of the orthorhombic transition itself. If
the latter is second order, one certainly expects the intermediate
phase to appear. However, if the transition is first-order, it can trigger a triclinic phase immediately, resulting in a first-order tetragonal-to-triclinic
transition. The latter seems to be the case for compositions closer
to the overdoped regime.

The remaining question left in this scenario is whether $\varepsilon_{3}$
condenses as the result of a lattice instability or as an indirect
consequence of the onset of the electronic nematic order parameter
$\psi$ that transforms as the $B_{1g}$ irrep. The observation of
CDW peaks at the (incommensurate) ordering vector $\mathbf{Q}_{1}=\left(q~0~0\right)$
suggest a possible charge-driven electronic nematic phase. Indeed,
denoting by $\Delta_{1}$ the CDW order parameter with ordering vector
$\mathbf{Q}_{1}$ and by $\Delta_{2}$ the CDW with ordering vector
$\mathbf{Q}_{2}=\left(0~q~0\right)$, CDW fluctuations can lead to
a nematic order parameter $\psi\propto\Delta_{1}^{2}-\Delta_{2}^{2}$. For our phenomenological analysis, however, the microscopic origin
of the nematic order parameter $\psi$ is not important.
The free energy expansion including the electronic nematic degrees
of freedom is given by:
\begin{eqnarray}
\mathcal{F}_{\mathrm{elast-nem}} & = & \frac{a_{s}}{2}\varepsilon_{3}^{2}+\frac{u_{s}}{4}\varepsilon_{3}^{4}+\frac{a_{n}}{2}\psi^{2}+\frac{u_{n}}{4}\psi^{4}-\lambda_{1}\psi\varepsilon_{3}\,.\label{eq:free_energy_b1g}
\end{eqnarray}
The last term is the symmetry allowed bilinear coupling between the
nematic order parameter and the orthorhombic distortion. Here, the
coefficient $a_{s}=C_{11}-C_{12}$, whereas the coefficient $a_{n}$
is proportional to the inverse of the bare nematic susceptibility,
$\chi_{\mathrm{nem}}^{-1}$. As usual, we write $a_{n}=c_{n}\left(T-T_{n,0}\right)$
and $a_{s}=c_{s}(T-T_{s,0})$, with $T_{n,0}$ the bare nematic transition
temperature and $T_{s,0}$, the bare structural transition temperature.
Hereafter, we set $c_{s}=c_{n}=1$ for simplicity. 

A state with non-zero $\varepsilon_{3}$ and $\psi$ can be the result
of a vanishing $a_{s}$ (lattice-driven) or of a vanishing $a_{n}$
(electronic-driven). To model the crossover from a lattice-driven
transition ($T_{s,0}<T_{n,0}$) to an electronic-driven transition
($T_{n,0}<T_{s,0}$), we define $T_{n,0}\equiv T_{0}\left(1+\delta\right)$
and $T_{s,0}\equiv T_{0}\left(1-\delta\right)$. Of course, $T_{0}$
and $\delta$ in principle depend on the doping composition $x$.
For our purposes, however, which is to contrast the behavior of the
elastoresistance in the lattice-driven and electronic-driven regimes,
we consider them independent variables and fix $T_{0}$ while varying
$\delta$. 

\begin{figure}
\centering \includegraphics[width=0.5\textwidth]{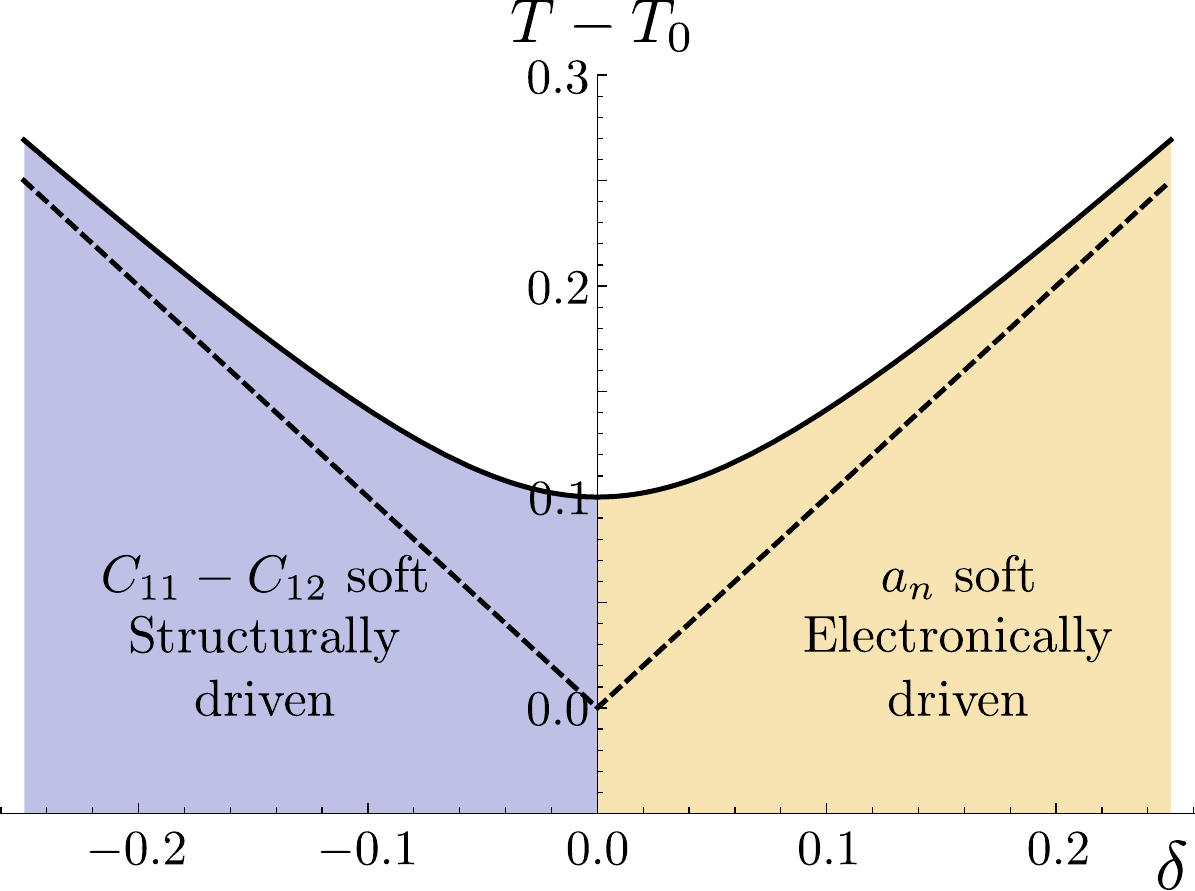}
\caption{\label{fig:schematic_phase_diagram} \textbf{Phase diagram for the coupled
structural-nematic transition.} The transition temperature $T_{\mathcal{N}}$
(full line) is affected by the finite coupling, $\lambda_{1}$, between
the structural and electronic degrees of freedom. Here we set $\lambda_{1}=0.10$.
In the absence of $\lambda_{1}$, the transition temperature is given
by the dashed line.}
\end{figure}

The resulting $(\delta,T)$ phase diagram is shown in Fig. \ref{fig:schematic_phase_diagram}.
Obviously, there is a single transition at $T_{\mathcal{N}}$ to a
state that lowers the tetragonal symmetry of the system down to orthorhombic,
given by the solid line. In the limit $\delta\ll-1$, the transition
temperature approaches $T_{s,0}$, whereas in the limit $\delta\gg1$,
the transition approaches $T_{n,0}$.

Our goal is to compute the quantity $\frac{\partial\psi}{\partial\varepsilon_{3}}$,
which is proportional to the elastoresistance $m_{12}-m_{11}$ in
the disordered state, for different values of $\delta$. A straightforward
calculation gives: 

\begin{equation}
\frac{\partial\psi}{\partial\varepsilon_{3}}=\frac{\lambda_{1}}{a_{n}+3u_{n}\phi^{2}}\,.
\end{equation}
\begin{figure}
\includegraphics[width=0.5\textwidth]{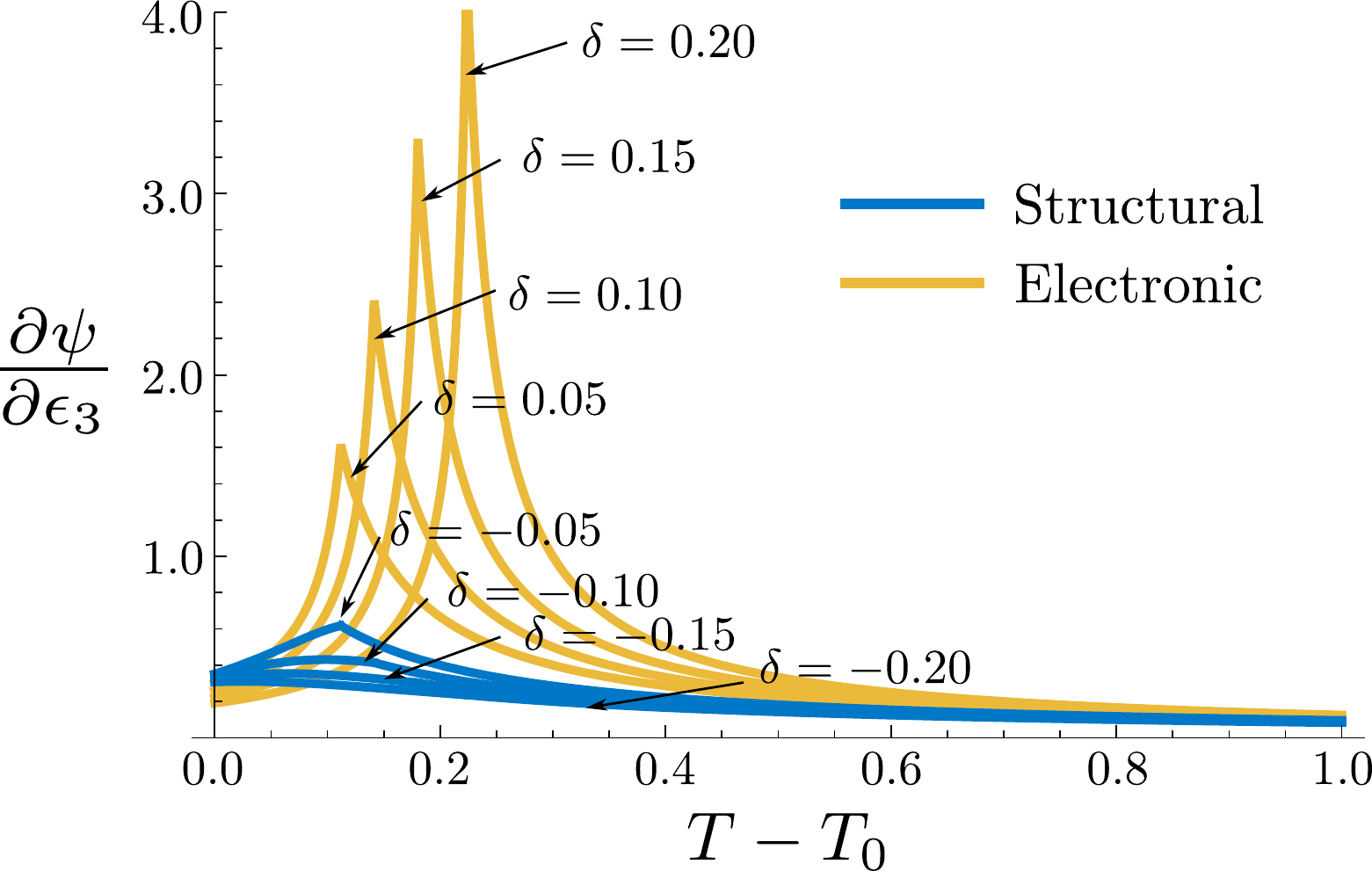} \caption{\label{fig:nem_susc} \textbf{Nematic susceptibility in both structurally-driven ($\delta<0$) and electronically-driven ($\delta>0$) regimes}. Here $\delta$ ranges from $\delta=-0.20$ to $\delta=0.20$. A sharp peak is evident in the electronically driven case. The parameters used were $\lambda_{1}=0.1$,
$u_{n}=1.0$, and $u_{s}=1.0$.}
\end{figure}
In Fig.~\ref{fig:nem_susc}, we plot $\frac{\partial\psi}{\partial\varepsilon_{3}}$
as a function of $T$ for different values of $\delta$, corresponding
to both the structurally driven ($\delta<0$) and electronically ($\delta>0$)
driven cases. In the structurally driven regime, there is no obvious
sign of a transition in $\frac{\partial\psi}{\partial\varepsilon_{3}}$
at $T_{\mathcal{N}}$, as this quantity remains nearly temperature-independent
or displays a mild peak. This is a consequence of the fact that the
bare nematic susceptibility, proportional to $1/a_{n}$, does not
diverge. We emphasize that the actual renormalized nematic susceptibility
does diverge, due to the contribution from the lattice degrees of
freedom. The behavior of $\frac{\partial\psi}{\partial\varepsilon_{3}}$
in the disordered state is similar to the experimentally observed
behavior of $m_{12}-m_{11}$ for the undoped composition BaNi$_{2}$As$_{2}$. 

In contrast, in the region of the phase diagram where the transition
is electronically-driven, $\frac{\partial\psi}{\partial\varepsilon_{3}}$
displays a clear peak at the transition temperature $T_{\mathcal{N}}$
and, within mean-field, a Curier-Weiss temperature dependence. This
behavior is reminiscent of that of $m_{12}-m_{11}$, in the disordered
phase, for underdoped and near optimally-doped Ba$_{1-x}$Sr$_{x}$Ni$_{2}$As$_{2}$.

\section{Supporting Experimental Data}

\subsection{X-Ray Diffraction and Charge Order}

\begin{figure}
    \centering
    \includegraphics[width=0.87\textwidth]{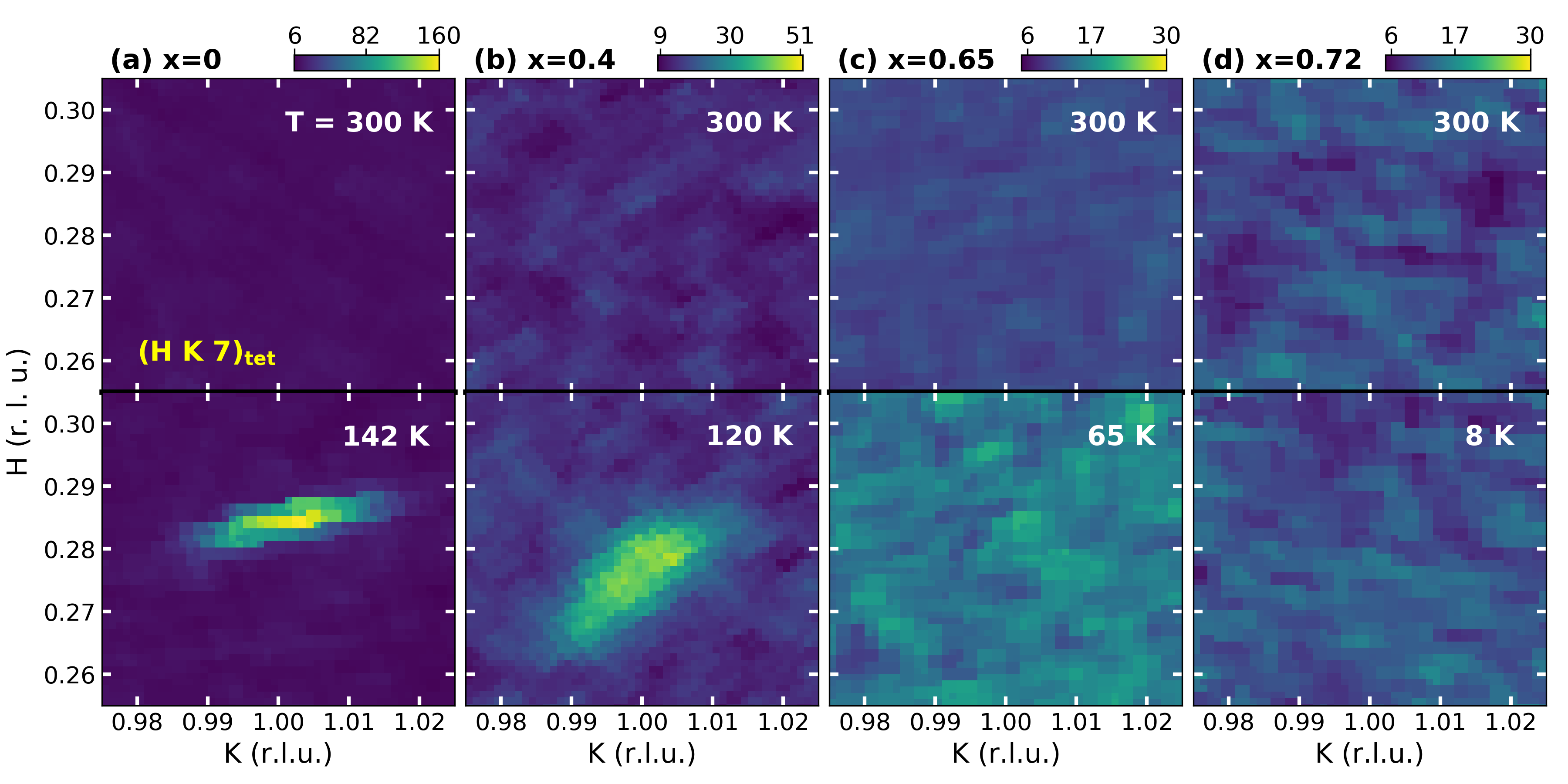}
    \caption{\textbf{I-CDW x-ray reflections in \BaSr~single crystals}. $(H~K)$ k-space mapping of x-ray scattering intensity in \BaSr~single crystals in the neighborhood of $(H~K~L)$ coordinates $(0.28~1~7)$. Data are presented at $x=0$ (a), 0.4 (b), 0.65 (c), and 0.72 (d). Top panel displays k-space mapping at 300 K, where no superstructure reflections are visible for all $x$. The lower panel displays data collected at temperatures just above the tetragonal-triclinic structural transition temperature for $x=0$, 0.4, and 0.65. Specimens of $x=0.72$ exhibit no structural transition, however, thus the lower panel of (d) instead displays the k-space intensity map collected at the base temperature of the experiment, 8 K.}
    \label{fig:CDW}
\end{figure}

Figure \ref{fig:CDW} displays $(H~K)$ mappings of reciprocal space, illustrating a representative I-CDW x-ray scattering peak in \BaSr~single crystals. I-CDW superstructure reflections are observed only in the tetragonal phase, at temperatures near the tetragonal-triclinic structural transition. The I-CDW abruptly vanishes upon transitioning to the low temperature triclinic structure, within which a new, commensurate superstructure is observed. Appearing with a $(0.28~0~0)$ periodicity in \BaNi, the superstructure wave vector appears to have a slight substitution dependence, appearing closer to $(0.27~0~0)$ in samples of $x=0.4$. Additionally, broader superstructure reflections are observed in $x=0.4$ samples than those in $x=0$. Estimates derived from peak full width at half maximum values suggest a correlation length of approximately 330 \AA~at $x=0.4$. I-CDW peak widths in $x=0$ samples, meanwhile, are narrower than the resolution of the experimental apparatus, placing a lower bound on the correlation length of nearly 1000 \AA. In both $x=0$ and $x=0.4$ crystals superstructure reflections are unidirectional, observed only along $(H~0~0)$ and not in the orthogonal $(0~K~0)$ direction, breaking $B_{1g}$ symmetry. Despite this symmetry breaking, no concurrent orthorhombic distortion is observed at the I-CDW onset within the resolution of the instrument. In more heavily substituted samples of Ba$_{0.35}$Sr$_{0.65}$Ni$_2$As$_2$ ($T_S=60$ K) the incommensurate peaks are not visible at 65 K (or any other measured temperature). While only a single representative peak is displayed in Fig. \ref{fig:CDW}, I-CDW reflections have been observed across several Brillioun zones in \BaNi~and Ba$_{0.6}$Sr$_{0.4}$Ni$_2$As$_2$, and are absent in the entirety of visible k-space at all temperatures in Ba$_{0.35}$Sr$_{0.65}$Ni$_2$As$_2$ and Ba$_{0.28}$Sr$_{0.72}$Ni$_2$As$_2$. All samples with an observable structural distortion ($x<x_c$) exhibit C-CDW peaks in the triclinic phase which are not displayed here.

\subsection{Structural Properties}

\begin{figure}
    \centering
    \includegraphics[width=0.57\textwidth]{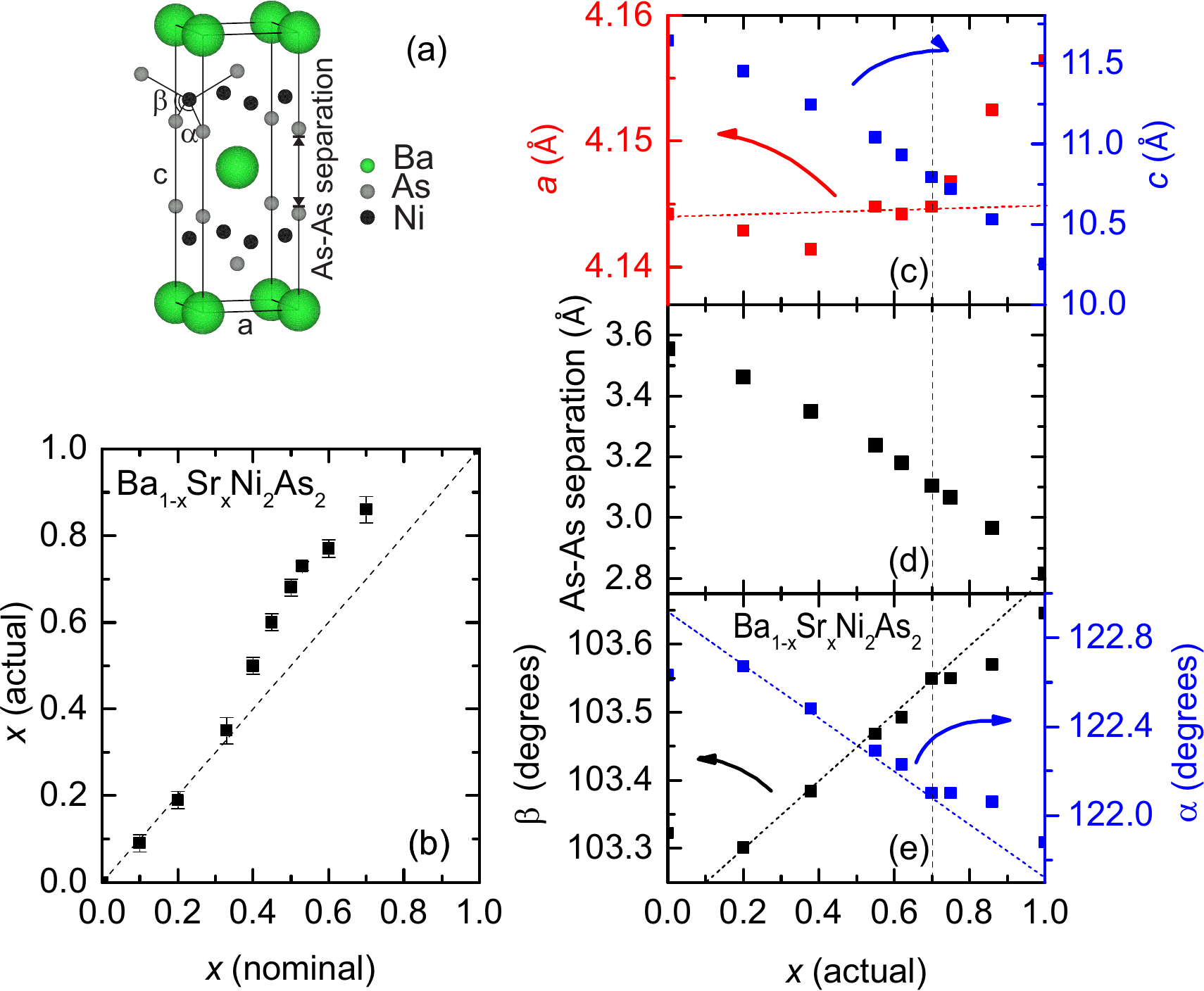}
    \caption{\textbf{Structural and chemical properties of \BaSr~single crystals.} (a) Crystal structure of tetragonal \BaNi. (b) Evolution of actual Sr content (determined via EDS) versus nominal value included in the growth. (c-e) $x$ dependence of \textit{a} and \textit{c} lattice parameters (c), As-As spacing (d), and tetrahedral bond angles (e) in \BaSr~single crystals. All data are collected at 250 K. The dashed line running vertically through all three panels denotes critical concentration, $x_c$, where the system undergoes a zero temperature structural phase transition.}
    \label{fig:XRAY}
\end{figure}

Structural parameters of \BaSr~single crystals collected from single crystal x-ray diffraction experiments at 250 K exhibit a subtle feature in the a-axis lattice parameter and tetrahedral bond angles near $x_c=0.7$ (Fig. \ref{fig:XRAY}). These features coincide with a sudden quench of the structural distortion, implying a change in interatomic bonding near $x_c$ may render the system inhospitable to the triclinic phase. Analogy to the iron based systems suggests this change may take the form of an interaction between interlayer arsenic atoms. Such an effect (as is observed in CaFe$_2$As$_2$ under pressure \cite{Kreyssig,Saha2012}) breaks no crystal symmetries and occurs as the interlayer As-As separation approaches 3 \AA, both of which are consistent with observations here. However, in the iron-based compounds interlayer As atoms will form a strong bond causing a dramatic collapse in the crystallographic c-axis, an abrupt reduction in As-As spacing, and an expansion of the transition metal-pnictogen layer, all of which are absent in the \BaSr~series. Thus, if an interaction between interlayer arsenic atoms is responsible for the structural anomaly in \BaSr~samples, it manifests in a fundamentally different manner than has been reported for iron-based compounds. 

\subsection{Resistivity, Magnetism and Specific Heat}

\begin{figure}
    \centering
    \includegraphics[width=0.67\textwidth]{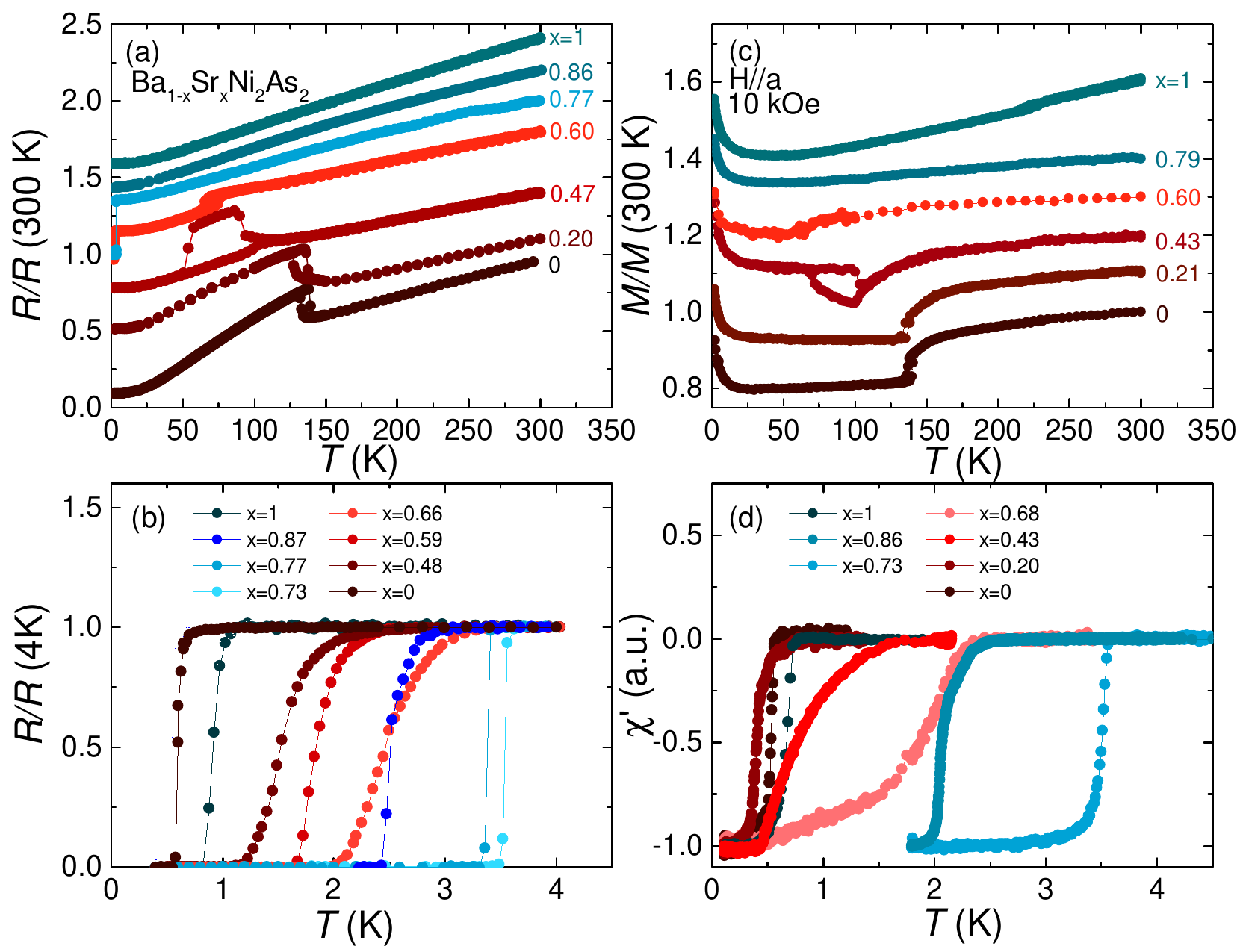}
    \caption{\textbf{Physical properties of \BaSr~single crystals.} Temperature dependent normalized resistivity data are presented for \BaSr~single crystals at high (a) and low (b) temperatures. Temperature hysteretic anomalies in (a) coincide with the tetragonal-triclinic distortion. (c) DC magnetization data collected at 10 kOe with $H\parallel a$ are presented. Sharp, temperature hysteretic anomalies at high temperature indicate structural transition, while divergence at low temperature likely originates from paramagnetic impurities. (d) Low temperature ac-magnetic susceptibility showing superconducting transitions. In all panels samples plotted in red (blue) are triclinic (tetragonal) at lowest temperatures.}
    \label{fig:RM}
\end{figure}

Figure \ref{fig:RM} displays basic magnetic and transport properties of \BaSr~single crystals. Temperature hysteretic anomalies indicate the onset of the first order tetragonal-triclinic structural distortion. The location of these anomalies was used to construct the phase diagram presented within the main text. Magnetization data similarly track the structural distortion, producing a sharp downward anomaly at the phase transition. 

The low temperature resistive and ac-magnetic signatures of superconductivity are presented in Fig. \ref{fig:RM}(b,d). While \BaSr~crystals in the range of $0.4<x<0.7$ exhibit transport and magnetic signatures of superconductivity at temperatures as high as 2.5 K, these signatures are significantly broadened in temperature compared to samples outside of this range. Furthermore, no heat capacity jump is observed at the magnetic and resistive superconducting transitions in these samples. Apparent superconducting signatures in transport and magnetic measurements are therefore believed to originate from superconducting filaments rather than a bulk superconducting phase, with bulk \Tc~remaining effectively constant for all crystals in the low temperature triclinic structure.

\begin{figure}
    \centering
    \includegraphics[width=0.57\textwidth]{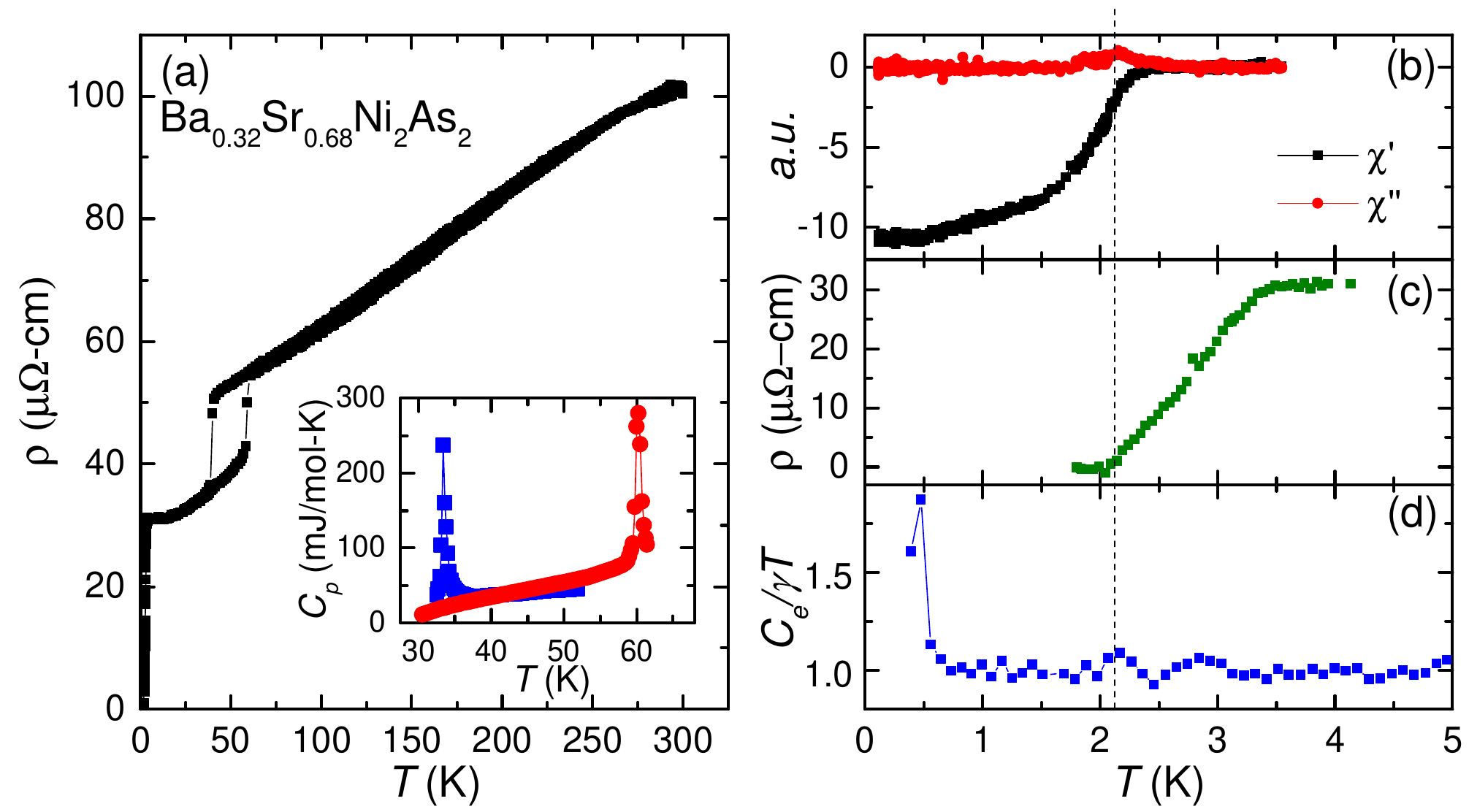}
\caption{\textbf{Physical properties of a single \BaSr~crystal of $x=0.68$}. (a) Transport data are taken between 1.8 and 300 K in a \BaSr~sample of $x=0.68$. Inset displays heat capacity data collected over the tetragonal-triclinic structural transition in the same crystal specimen. Anomalies associated with the structural transition are observed in heat capacity, showing it is of bulk origin and is strongly first order. (b-d) Magnetization (b), transport (c), and heat capacity (d) superconducting transitions observed again in the same single crystal measured in panel (a). While magnetization and transport signatures suggest a \Tc~of approximately 2 K, a heat capacity anomaly is only observed near 0.5 K, indicating the bulk superconducting \Tc~is virtually identical to the \BaNi~end-member.}
    \label{fig:115}
\end{figure}

That \Tc~enhancement in undersubstituted samples is filamentary in nature is observed clearly in resistivity, heat capacity, and magnetization measurements collected from a single representative underdoped $x=0.68$ crystal [Fig. \ref{fig:115}(a-d)]. Heat capacity and transport measurements confirm a hysteretic tetragonal to triclinic structural transition [Fig. \ref{fig:115}(a)]. Data taken below 5 K exhibit resistive and magnetic superconducting signatures that are broadened, and occur at approximately 2 K. Heat capacity data, however, shows a sharp anomaly at a much lower temperature, 0.5 K [Fig. \ref{fig:115}(d)]. This behavior has been observed across a number of samples, and no crystals have been found where both the triclinic structural distortion and enhanced superconducting \Tc~compared with the \BaNi~end-member are present.  

\begin{figure}
    \centering
    \includegraphics[width=0.47\textwidth]{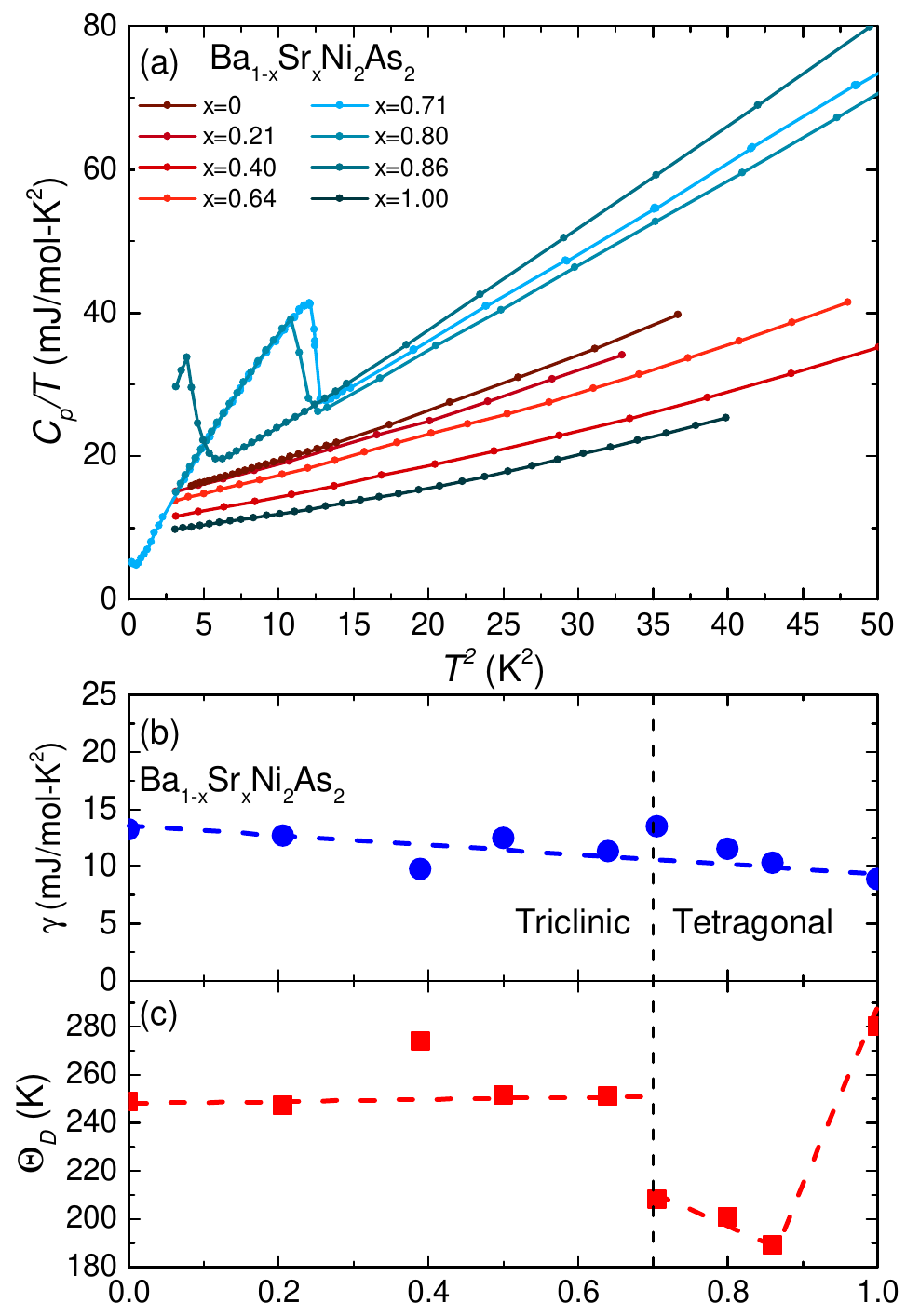}
    \caption{\textbf{Heat capacity data for \BaSr~series.} (a) $C_p/T$ vs $T^2$ values are presented for many $x$ in the \BaSr series. Data denoted with red (blue) symbols indicate that samples of this concentration favor a triclinic (tetragonal) structure at lowest temperature. Evolution of the Sommerfeld coefficient (b) and Debye temperature (c) are included as a function of $x$. Dashed line running vertically through $x_c=0.7$ denotes the critical Sr concentration when the system transitions from triclinic to tetragonal crystal symmetries at lowest temperature.}
    \label{fig:CP}
\end{figure}

Additionally, heat capacity measurements in \BaSr~crystals show a clear discontinuity in Debye temperature when crossing the zero temperature structural phase boundary [Fig. \ref{fig:CP}(c)]. In the oversubstituted samples, the Debye temperature remains effectively constant between $x=0.7$ and $x=0.87$ before increasing in the SrNi$_2$As$_2$ end member. The Sommerfeld coefficient ($\gamma$) exhibits a minimal $x$ dependence across the phase diagram, decreasing modestly between $x=0$ and $x=1$. Both Debye temperature and Sommerfeld coefficient are extracted from heat capacity data using an Einstein-Debye model.

\subsection{Elastoresistivity}

\begin{figure}
    \centering
    \includegraphics[width=0.47\textwidth]{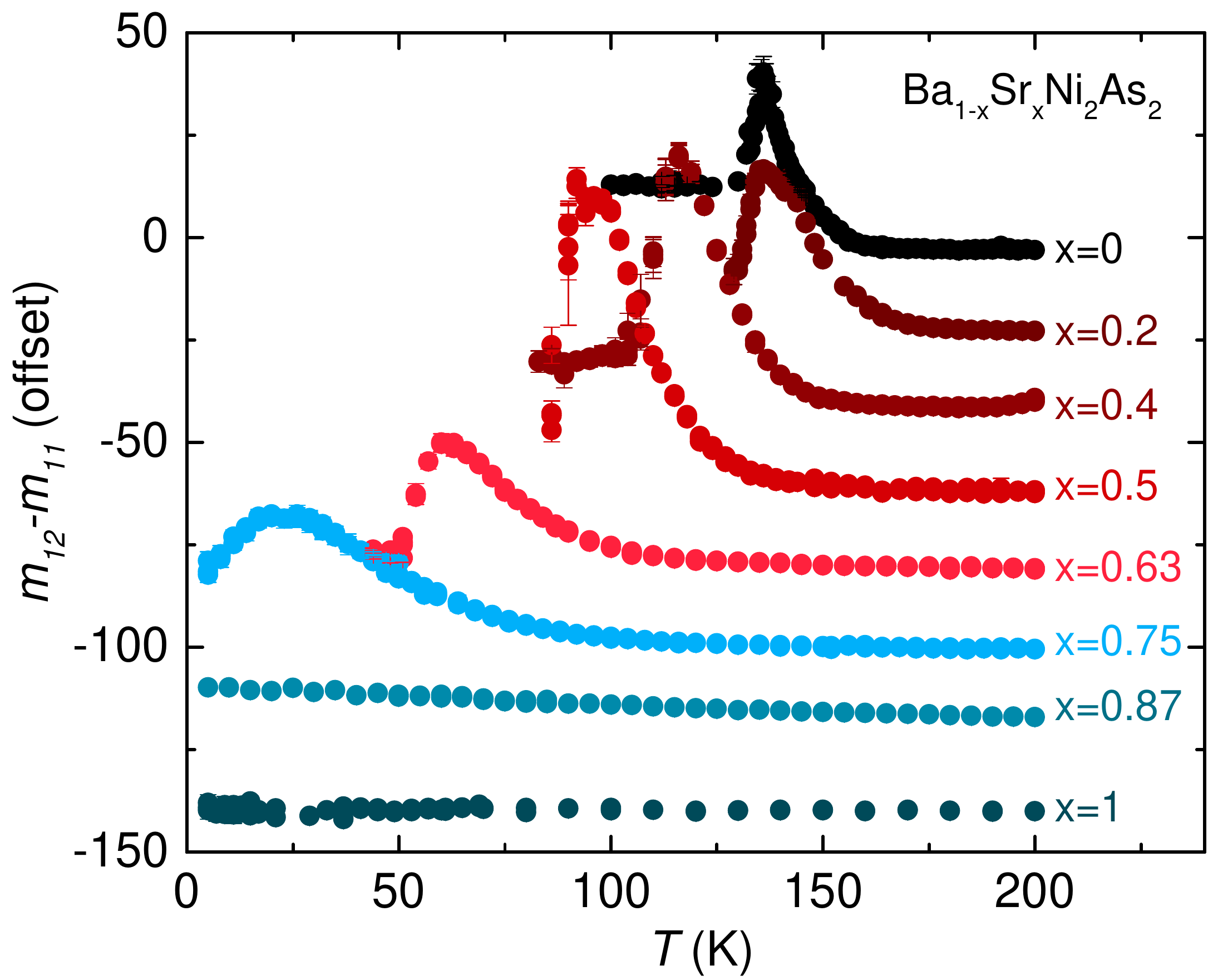}
    \caption{\textbf{$m_{12}-m_{11}$ elastoresistivity of \BaSr~single crystals}. $B_{1g}$ symmetric elastoresistivity for \BaSr~single crystals of select $x$ are presented. Data denoted with red (blue) symbols indicate that samples of this concentration favor a triclinic (tetragonal) structure at lowest temperature. Values presented here were used to generate the contour plot presented in the main text.}
    \label{fig:ALL}
\end{figure}

\begin{figure}
    \centering
    \includegraphics[width=0.47\textwidth]{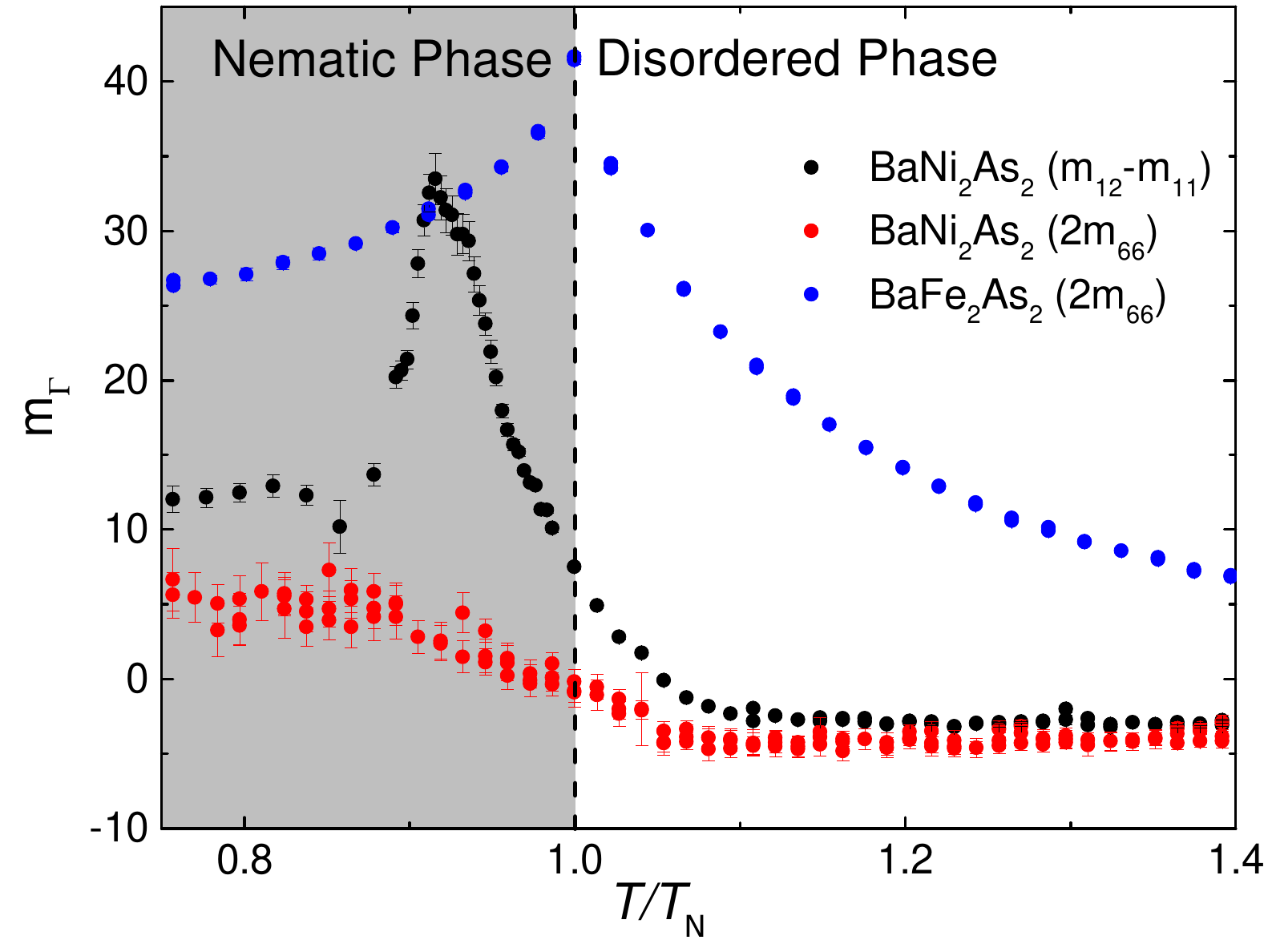}
    \caption{{\bf Elastoresistivity comparison, \BaNi~and \BaFe.}
 Elastoresistivity is plotted versus normalized temperature (normalized to the nematic transition temperature, $T_\mathcal{N}$) for \BaNi~and \BaFe~single crystals. Data are provided below $T_\mathcal{N}$, though, due to the structural change and, in the case of \BaNi, onset of elastoresistive hysteresis, these data are no longer well defined symmetry isolated elastoresistive components. }
    \label{fig:FENI}
\end{figure}

\begin{figure}
    \centering
    \includegraphics[width=0.47\textwidth]{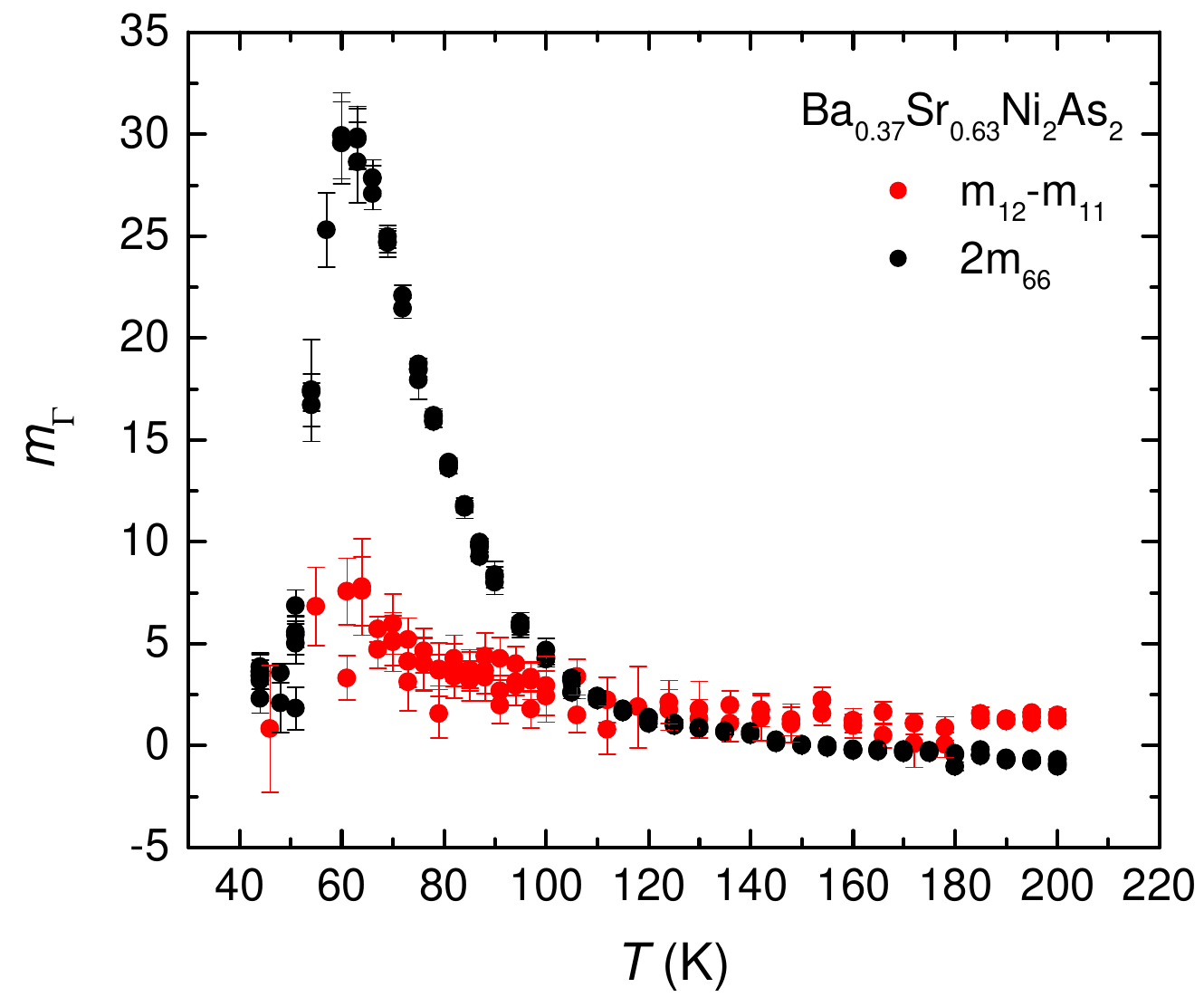}
    \caption{\textbf{Elastoresistivity in undersubstituted \BaSr}. $B_{1g}$ and $B_{2g}$ symmetric nematic susceptibility for samples of $x=0.63$. Data exhibits a Curie-Weiss reminiscent temperature divergence in the $B_{1g}$ channel, while $B_{2g}$ symmetric susceptibility does not diverge.}
    \label{fig:115N}
\end{figure}

Figure \ref{fig:ALL} displays the measured elastoresistivity for \BaSr~single crystals of select $x$. Data are staggered by a constant offset. Figure \ref{fig:FENI} displays the measured elastoresistivity for both \BaNi~and \BaFe~single crystals. \BaNi~$B_{2g}$ symmetric elastoresistivity exhibits almost no response to the onset of nematic order. The $B_{1g}$ symmetric elastoresistivity ($m_{12}-m_{11}$) similarly does not diverge approaching the nematic transition in unsubstituted \BaNi, a fact made more obvious through comparison to the clear divergence seen in \BaFe. Meanwhile, in samples of $x=0.63$, a clear divergence in $B_{1g}$ symmetric elastoresistivity is observed approaching the structural transition, reminiscent of the $B_{2g}$ response in \BaFe. The $m_{66}$ elastoresistivity coefficient in $x=0.63$ shows virtually no temperature dependence, however, indicating that $B_{2g}$ symmetric nematic fluctuations do not strengthen significantly in the \BaSr~series with increasing $x$.

\begin{figure}
    \centering
    \includegraphics[width=0.47\textwidth]{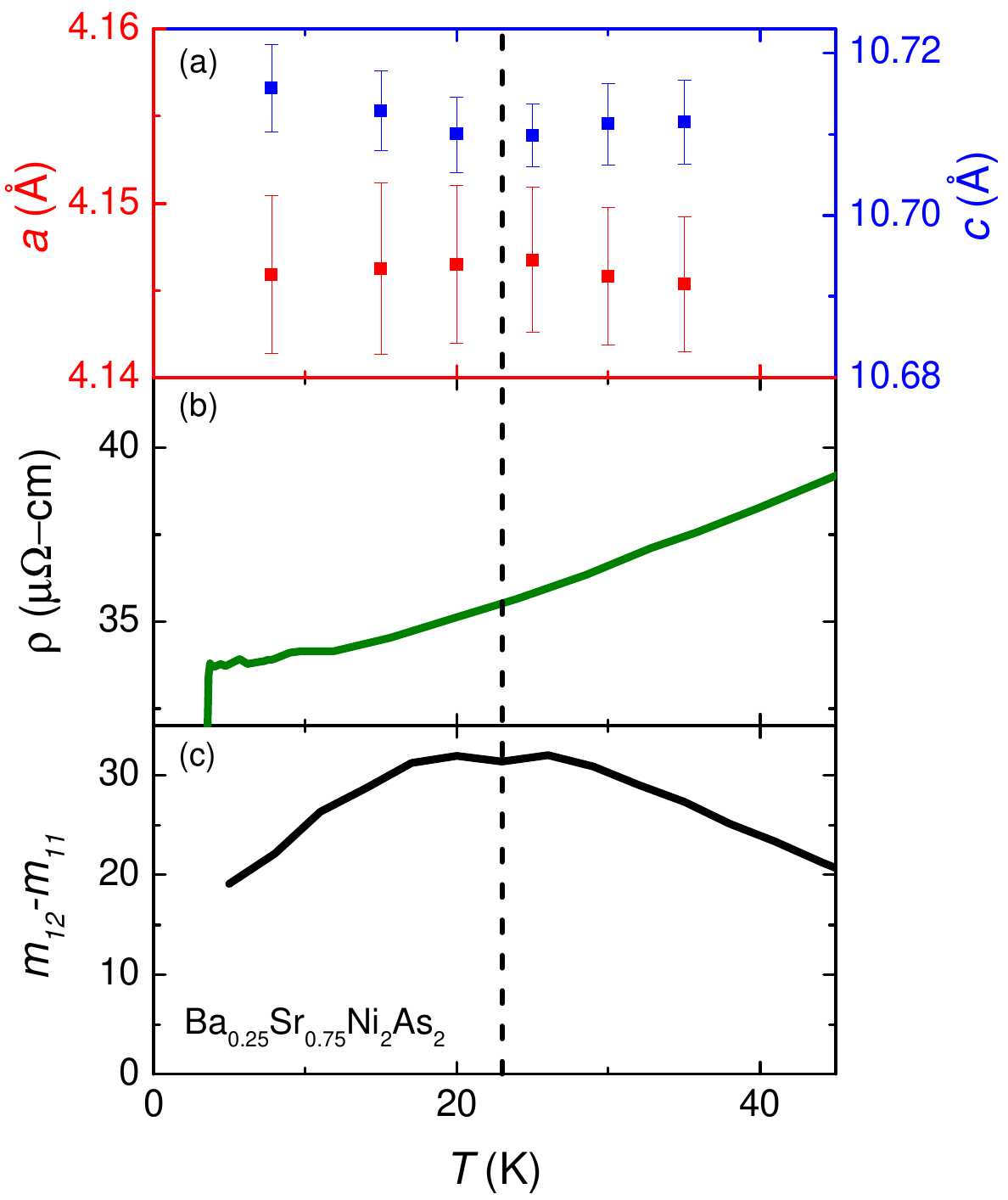}
    \caption{\textbf{Physical properties of optimally substituted \BaSr}. (a-c) Evolution of \textit{a} and \textit{c} crystallographic lattice parameters (a), resistivity (b), and elastoresistivity (c) as a function of temperature in optimally substituted \BaSr. While elastoresistivity displays a peak, no sign of an apparent phase transition is observed in either transport or structural measurements.}
    \label{fig:OPT}
\end{figure}

\BaSr~specimens of $x=0.75$ show a large $m_{12}-m_{11}$ elastoresistive response, reaching a maximum value of nearly 35. X-ray diffraction measurements show no structural distortion or superstructure reflections between room temperature and 8 K, while transport measurements do not indicate of a phase transition [Fig. \ref{fig:OPT}(a,b)]. Despite the absence of any apparent transition $m_{12}-m_{11}$ coefficients display a clear peak and subsequent downturn near 25 K [Fig. \ref{fig:OPT}(c)].

\end{document}